\documentclass[iop]{emulateapj}

\usepackage{natbib}		
\usepackage{amsmath}		
\usepackage{graphicx}		
\usepackage{hyperref}		

\newcommand {\eq}[2]		
{
   \begin{equation}  \label{#1}
   \begin{aligned}
      #2
   \end{aligned}
   \end{equation}
}

\newcommand {\tab}			
{
   \;\;\;\;\;
}

\begin{document}

\title{
   Evolution of a Pulsar Wind Nebula within a Composite Supernova Remnant
}





\author {
  Christopher Kolb \altaffilmark{1},
  John Blondin     \altaffilmark{1},
  Patrick Slane    \altaffilmark{2},
  Tea Temim        \altaffilmark{3}
}

\altaffiltext{1}{North Carolina State University}
\altaffiltext{2}{Harvard-Smithsonian Center for Astrophysics}
\altaffiltext{3}{Space Telescope Science Institute}

\date{\today}

\begin{abstract}
The interaction between a pulsar wind nebula (PWN) and its host supernova remnant can produce a vast array of observable structures. Asymmetry present within these structures derives from the complexity of the composite system, where many factors take turns playing a dominating hand throughout the stages of composite SNR evolution. Of particular interest are systems characterized by blastwave expansion within a non-uniform interstellar medium (ISM), which contain an active pulsar having a substantial `kick' velocity (upward of 300 $\text{km s}^{-1}$), as these systems tend to produce complex morphologies. We present a numerical model which employs these and several other factors in an effort to generate asymmetry similar to that seen in various x-ray and radio observations. We find the main parameters driving structure are ISM uniformity and total pulsar spin-down energy, with secondary contributions from factors such as pulsar trajectory and initial spin-down luminosity. We also investigate dynamics behind PWN `tails', which may form to link active pulsars to a crushed, relic nebula as the reverse shock passes. We find that the direction of such tails are not good indicators of pulsar motion, but direction does reveal flow of ejecta created by the passage of a reverse shock.
\end{abstract}

\section{Introduction}

The core collapse of a massive star may produce a magnetized, rapidly rotating neutron star (pulsar), capable of emitting a wind of relativistic, charged particles with luminosity on order of $10^{38}\ \text{erg s}^{-1}$. Emission of this wind works rapidly to inflate a bubble of shocked gas surrounding the compact object, deemed a pulsar wind nebula (PWN). The composite supernova remnant (SNR) is defined by the evolution of this PWN within the interior region of the collapsed star. While the majority of these objects retain spherical symmetry during early-stage development \citep[e.g.][]{2000ApJSlane}, the composite system is of particular interest as its relic PWN tends to form complex structure during late-stage evolution. Evidence of this complex morphology may be observed through features such as asymmetric nonthermal emission 
\citep{
2007ApJPark, 		
2013ApJTemim,		
2014ApJLBorkowski},	
kicks 
\citep{
1998ApJWang,		
2009ApJHales},		
and unique PWN structure 
\citep{
2012ApJSlane,		
2001ApJKaspi}.		
In most cases, significant asymmetry is attributed to a passage of the reverse shock, where shocked SNR ejecta violently mixes with PWN material.

Motivation for this paper is found in the unique structure present within the composite SNR G327.1-1.1. Observational data from the \textit{Chandra} X-ray Observatory reveals a neutron star within the remnant, located east of the shell's geometric center and traveling northward with an inferred speed of 400 $\text{km s}^{-1}$. To the pulsar's southeast exists a large, spherical radio nebula which is connected to the compact source through a narrow tail-like structure. The age of the system is believed to be approximately 17 kyr, with current analysis indicating the reverse shock has likely passed through the system and separated the pulsar from the relic nebula \citep{2015ApJTemim}. Further, dominant X-ray emission in the 2.58-7.0 keV band reveals a unique prong-like structure extending from the emission source toward the north-west, into oncoming residual flow created by the passage of the reverse shock.


Recent works investigating composite SNR morphology tend to assume a large degree of symmetry while employing both analytical and numerical methods. An analytical approach generally requires consideration of spherically symmetric evolution within a uniform interstellar medium (ISM) \citep[e.g.][]{1984ApJReynolds, 1995PyRepMcKee, 2001A&AVDS}, as systems which deviate strongly from symmetry prove difficult to tackle. Numerical models which deviate from spherical symmetry have generally focused on developing shell asymmetry through interaction with a non-uniform ISM.
Such models include the interaction between a two-dimensional (2D) blastwave and an interstellar cloud/ring 
\citep{
1999ApJJun,		
1997ApJBorkowski}, 	
2D composite systems expanding within a shelf-like density gradient 
\citep{
2001ApJBlondin,		
2008A&AFerreira},	
and three-dimensional (3D) blastwave propagation into linear and exponential ISM density gradients 
\citep{
1999A&AHnatyk}.		

This paper considers in detail two numerical models which have been chosen due to their predictable anisotropic nature. The first model \citep[][hereafter, vSDK]{2004A&AVDS} describes a spherical SNR expanding within a uniform ISM. The model introduces asymmetry through a pulsar, enclosed within a PWN, which is traveling at a constant rate due to some initial `kick' velocity. The high pulsar velocity used within this paper allows the object to reach one side of the reverse shock wall shortly after the system has entered the PWN crush phase. The model demonstrates the ability of a young pulsar's high-luminosity particle emission to form a bow shock-like structure around the compact object, allowing it to separate from the crushed PWN. The simulation develops a wide tail leading from the pulsar toward the bulk of the PWN nebula, with an axis of symmetry about the pulsar's motion.

The second model \citep[][hereafter, BCF]{2001ApJBlondin} consists of a blastwave expanding into a shelf-like non-uniform ISM density. The blastwave contains a motionless pulsar / PWN at its center, which emits energy through a time-dependent spin-down luminosity. The density shelf within this model produces an anisotropic mechanism, alternate to vSDK, which allows the pulsar to interact exclusively with one side of the collapsing reverse shock. The difference in blastwave propagation speeds within the two regions has an immediate impact on the reverse shock radius, creating both an interaction timescale similar to that seen in vSDK and a tendency for the high-density region's reverse shock to arrive back at the origin much earlier. An axis of symmetry develops perpendicular to the shelf.

The crush phase seen in BCF is significantly more violent than that seen in vSDK. This likely results from BCF's addition of a time-dependent correction to spin-down luminosity, which limits energy available to halt the collapsing reverse shock. Blondin et al. find that the crush is halted at some volume predicted by this total spin-down energy input, and the pulsar is located near the edge of the crushed PWN as the system enters a secondary slow expansion phase.

Here we combine these two models in order to develop an asymmetric model of the composite SNR system. We attempt to retain some degree of predictability for anisotropic elements by assuming a linear combination of true pulsar motion from the vSDK model and an apparent pulsar motion due to the ISM gradient from the BCF model. We will also use this model to investigate the effect true pulsar motion has on both PWN size and location post-PWN crushing.

\section{Methods}

The composite SNR model employed in this paper is primarily derived from BCF. The system assumes a spherical, self-similar driven wave solution to represent the blastwave \cite{1982ApJChevalier}, which is placed into an ISM density gradient outlined below. A freely expanding adiabatic gas bubble is then placed near the object's center to represent a young PWN (for further discussion, refer to BCF). All parameters are scaled to agree on a startup time of fifty years after the supernova event.

\subsection{Asymmetric Model}

Two mechanisms are used to generate asymmetric evolution within the model: pulsar `kick' velocity (see vSDK) and strength of the ISM density gradient. During uniform free expansion of the unshocked SNR interior, the kick acts to translate both the pulsar and the young PWN away from the explosion center at a velocity $\mathbf{v}_\text{psr}$. The trajectory of this motion is defined by an angle $\psi_\text{kick}$, measured with respect to the high-density region of the ISM. The imposed gradient affects both the propagation speed of the outer blastwave and the buildup of pressure behind it. The gradient itself is described by the equation

\eq{rho} 
{
   \rho(z) = 
   \rho_0 \left[ 1 + \frac{2-x}{x} \tanh\left( \frac{z}{H} \right) \right], \tab
   z = r\cos\theta 
}

\noindent
where $\rho_0$ is the average ambient density, $\theta = 0$ sets the gradient's highest-density point, $x$ is a scaling factor used to produce a high-to-low asymptotic density ratio of $(x-1)^{-1}$ as $z \to \infty$, and $H$ is a characteristic length-scale describing the hyperbolic tangent's `linear' region. We have quantified the gradient strength produced by equation \eqref{rho} in terms of the approximate change in particle density per unit length

\eq{gradrho} 
{
   \nabla\rho = \frac{\rho_\text{max} - \rho_\text{min}}{2Hm_p},
}

\noindent
with $H$ in pc, giving units of $\text{cm}^{-3}\ \text{pc}^{-1}$.

Shell geometry remains remarkably spherical while the blastwave is contained within the linear portion of the ISM density gradient, becoming slightly elliptical due to a differential propagation through the ISM \citep{1999A&AHnatyk}. Once the blastwave reaches the edge of this linear region at a radius of $R_\text{snr} \sim H/2$, the portion propagating into the high-density portion of the ISM will begin to decelerate much more rapidly than its low-density counterpart, and the shell will take on an egg-like shape (seen in various figures in BCF). Once both the high-density and low-density regions of the blastwave reach the flat portion of the hyperbolic tangent gradient, the shell will begin to deform into a dual-hemispherical shape. (We define this type of gradient to be `shelf-like,' where the blastwave diameter is $\gg H$ such that $\nabla\rho \to 0$ on the scale of the remnant.) In each instance, a differential buildup of pressure behind the blastwave will drive a similar deformation of the reverse shock.

\begin{figure}[]
   \centering
   \includegraphics[width=3in]
      {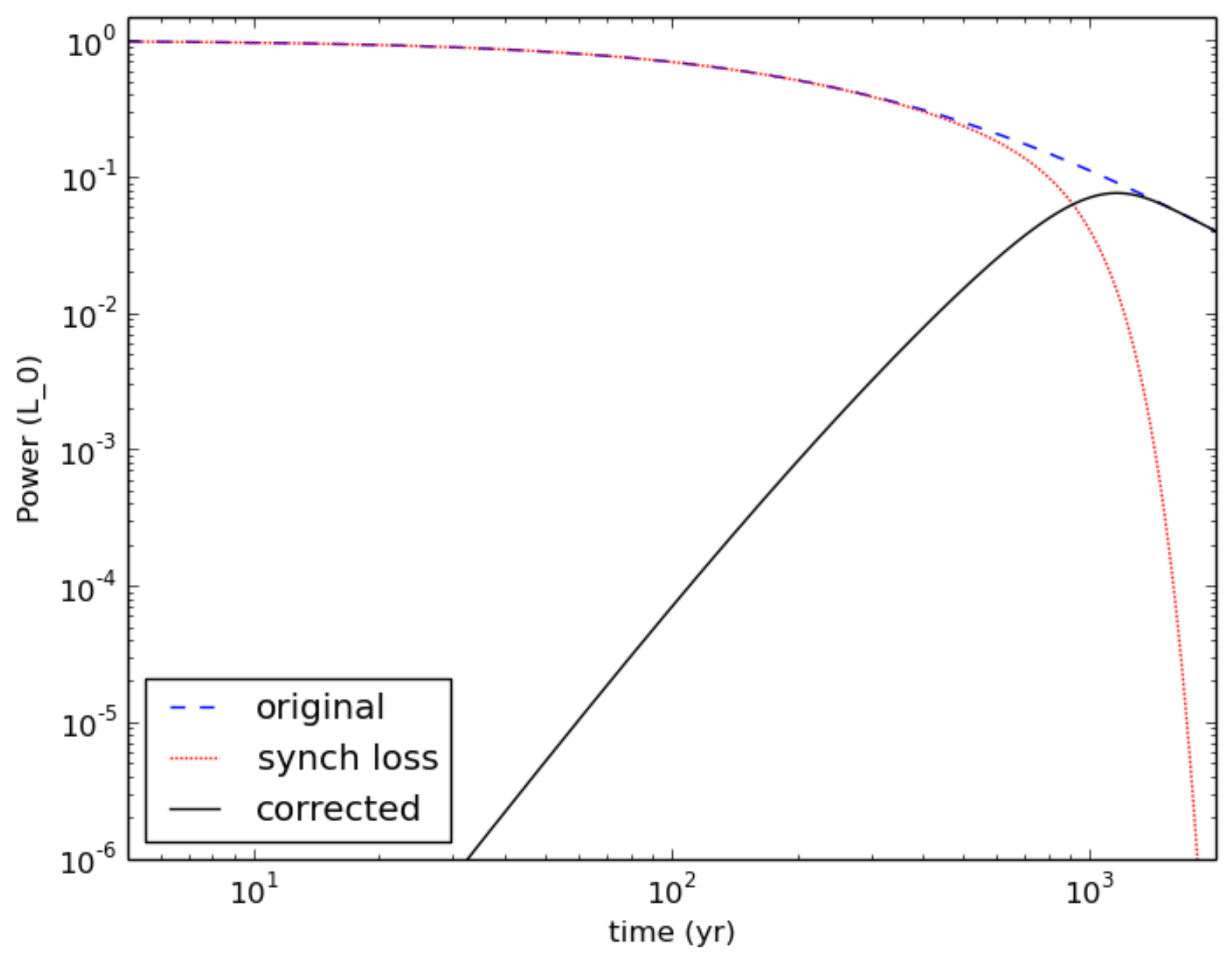}
   \caption
      {Visualization of the pulsar spin-down luminosity using the correction for synchrotron losses shown in Equation \eqref{Edot} ($\tau = 500\ \text{yr}$). The original spin-down luminosity term $L_p$ from BCF is represented by a dashed line (blue), while the synchrotron-limited energy term $\dot E$ from Equation \eqref{Edot} is represented by a solid line (black). Synchrotron loss is represented by a dotted line (red) and closely follows results from \cite{2009ApJGelfand}. We note that this approximation only corrects for power provided to the relativistic particles and does not consider pressure from the magnetic field.	\label{effLum}}
\end{figure}

\begin{table*}[t]
  \tiny		
  \begin{center}
  \begin{tabular}{c}	
  \textbf{Model Parameters}  \\ [0.25em]
  \hline
  \hline \\
  \begin{tabular}{c | c c c c c c c c}

   			&
   Explosion		&
   Ejecta mass 		&
   Spin-down 		&
   Spin-down 		&
   Pulsar kick 		&
   Pulsar kick 		&
   Asymptotic 		&
   Characteristic	  \\ 

   			&
   energy 		&
			&  
   luminosity		&
   timescale		& 
   velocity		& 
   direction 		&
   density ratio	& 
   length scale		  \\ [0.2em]

   			&
   ($10^{51}$ erg) 	&
   ($M_\odot$) 		&
   ($10^{38} \text{erg s}^{-1}$) &
   (kyr) 		&
   (km s$^{-1}$) 	&
   (\textit{radians})	&
    			&
   ($10^{18}$ cm) 	  \\ [0.1em]

   Model		&
   $E_{51}$		&
   $M_\text{ej}$	&
   $L_\text{p}(0)$	& 
   $\tau$		& 
   $v_\text{psr}$	& 
   $\psi_\text{kick}$	&
   $(x-1)^{-1}$		& 
   $H$			  \\ 
   \hline

    A 			&
    0.5 		&
    4.5 		&
    2.8 		&
    2.0 		&
    350 		&
    $3\pi/4$ 		&
    12.5 		&
    1.6 		  \\ [0.1em]

    B 			&
    1.0 		&
    4.0 		&
    2.8 		&
    2.0 		&
    350 		&
    $\pi/4$ 		&
    20.0 		&
    30 			  \\ [0.1em]

    C 			&
    1.0 		&
    3.5 		&
    1.0 		&
    0.75		&
    300 		&
    $3\pi/8$ 		&
    10.0 		&
    3.0 		  \\ [0.1em]

    D 			&
    1.0 		&
    4.5 		&
    2.8 		&
    1.5 		&
    500 		&
    *\textit{variable} 	&
    12.5 		&
    12  		  \\ [0.1em]

    E 			&
    1.0 		&
    4.5 		&
    2.8 		&
    1.5 		&
    500 		&
    $7\pi/8$ 		&
    12.5 		&
    12			  \\ [0.1em]

  \end{tabular}
  \end{tabular}
  \end{center} 
\caption{	\label{parameters}}
\end{table*}

As total energy available plays a critical role in determining PWN size during late-stage morphology (BCF), the original model is further modified to consider energy lost to the PWN through synchrotron emission. This radiative emission has been shown to dominate over adiabatic energy loss up to a time approximately twice the spin-down timescale $\tau$, after which it falls off sharply \citep{2009ApJGelfand}. As we are primarily interested in late-stage morphology, this energy loss is modeled by appending an inverse-decay term to the original pulsar spin-down luminosity, scaled to a cutoff time of $t \sim 2\tau$

\eq{Edot} 
{
   \dot{E}(t) =
   L_p(t) \left\{ 1 - \text{exp}\left[ -\left(\frac{t}{2\tau}\right)^4 \right] \right\},
}

\noindent
where $L_p(t)$ is the spin-down luminosity defined in BCF as

\eq{Lp} 
{
   L_p(t) = L_0\left( 1 + \frac{t}{\tau} \right)^{-\frac{n+1}{n-1}}.
}

\noindent
(See Fig. \ref{effLum} for a comparison between the original spin-down luminosity $L_p(t)$ and this corrected term $\dot{E}(t)$.) This decay term accounts for radiative loss by limiting the energy flux onto the grid rather than altering the energy of existing grid elements. Such treatment is expected to slow propagation of the PWN shock during its initial supersonic expansion phase, as particles entering the grid during early times are given significantly less momentum than what is expected from a method calculating radiative losses on a zone-by-zone basis. The exponential in Equation 3 is chosen as it closely approximates total energy loss through synchrotron emission as presented in \cite{2009ApJGelfand} (one may compare results therein to the corrected spin-down luminosity in Fig. \ref{effLum}). We note that \cite{2009ApJGelfand} assumes a single power-law energy distribution while evidence for a much more complicated injection specturm exists \citep{2008ApJLSlane}. For the case of a broken power-law, there is typically more energy in particles above the break than below, and as power is proportional to energy squared, particles below the break are of little concern for synchrotron losses. Since these effects are not particularly large, we feel our approach provides a reasonable estimate of energy available to the PWN during the later stages of morphology.

To investigate general directional dependence within the relic PWN, an age tracking variable has also been implemented. The variable is input at the inner boundary and is designed to represent time since PWN gas has been released by the pulsar (and promptly shocked). This method provides insight into where pockets of aged material develop within composite systems and allows for direct observation of gas flow within the PWN during evolution.

\subsection{Numerical Approach}

The simulation is evolved using the grid-based hydrodynamics code VH-1, which utilizes the PPMLR method \citep{1984JCPC&W} to accurately resolve shock propagation. The 2D models presented here are computed on a cylindrical grid while assuming spherical convergence (i.e., an equatorial slice from a sphere) and utilize 576 radial zones and 576 angular zones, providing an angular resolution of 0.625 degrees. 3D models use a Yin-Yang grid \citep{2010A&AWong} consisting of two separate, overlapping spherical grids each containing 480 radial zones, 480 azimuthal zones, and 160 polar zones, for an angular resolution of $0.75$ degrees.

\begin{figure}[h]
   \centering
   \includegraphics[width=3in]
      {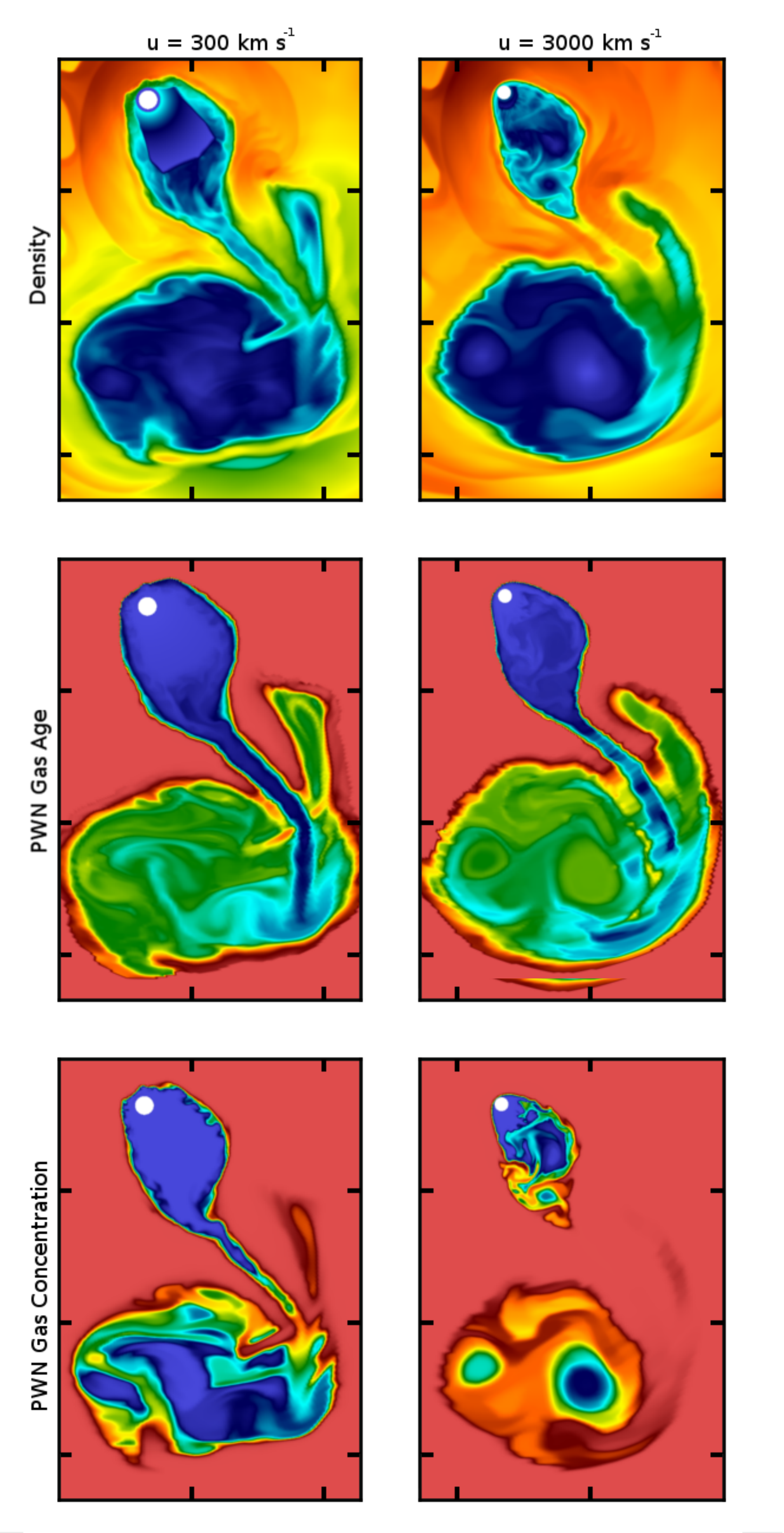}
   \caption
      {Comparison of asymmetric PWN morphology for an arbitrary system with a pulsar wind speed of $v_w = 3 \times 10^2\ \text{km s}^{-1}$ (this paper) and a faster speed of $v_w = 3 \times 10^3\ \text{km s}^{-1}$. \textit{Density} is represented on a log-scale with (blue) indicating the lowest and (red) indicating the highest values. \textit{PWN Gas Age} is represented with (blue) indicating most-recent emission from the pulsar and (red) indicating gas at approximately the age of the simulation. \textit{PWN Gas Concentration} is represented with (blue) indicating highest concentration of PWN gas and (red) indicating lowest concentration of PWN gas. Morphology remains consistent across wind speeds as evident by both the density and gas age plots. Further, the tail is seen in the PWN gas age plot to retain its shape despite a significant decrease in PWN gas concentration.   \label{windcomp}}
\end{figure}

Initial conditions for the self-similar driven wave are generated from the solution described in \cite{1982ApJChevalier} and scaled using model-dependent parameters. The PWN is similarly generated using a wind-blown bubble solution for an adiabatic gas with a mass-loss rate of $\dot{M} = 2\dot{E}/v_w^2$, where the velocity of the emitted wind has been scaled to a speed of $3 \times 10^2\ \text{km s}^{-1}$ in an effort to limit computational time. Although this scaling method leads to an artificially large PWN mass, the additional inertia has little effect on PWN morphology. This is evident in Figure \ref{windcomp}, where a comparison between the chosen speed of $3 \times 10^2\ \text{km s}^{-1}$ and a more realistic speed of $3 \times 10^3\ \text{km s}^{-1}$ shows consistent morphology and gas flow despite the large difference in PWN mass.

\begin{figure*}[t]
   \centering
   \includegraphics[width=\textwidth]
      {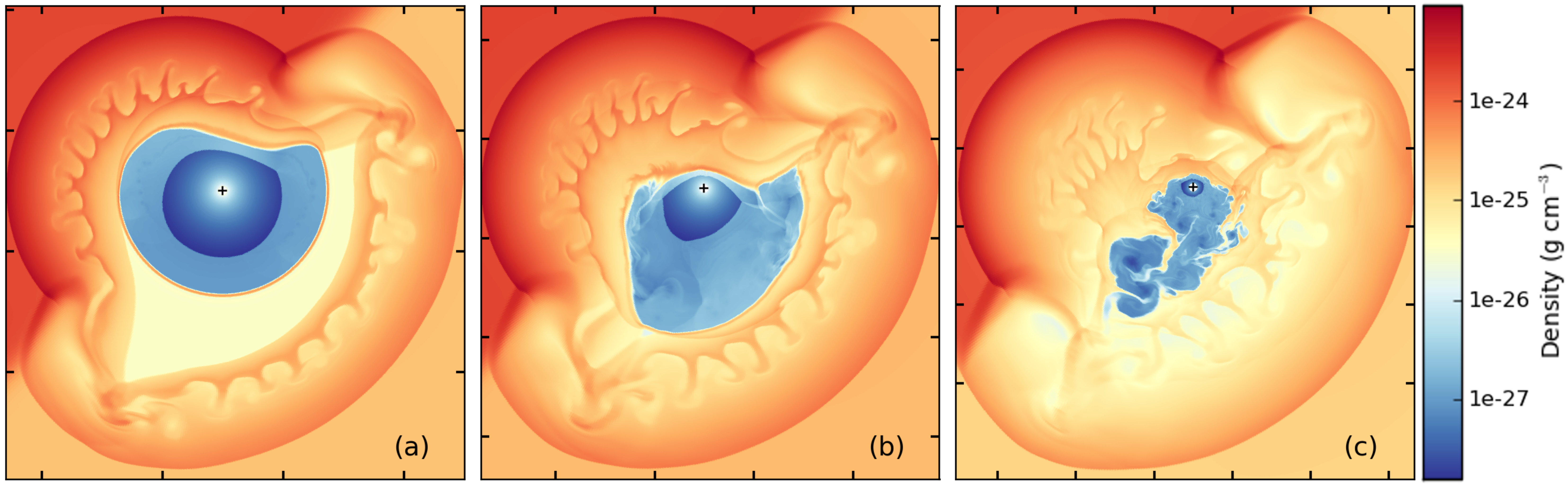}
   \caption
      {General evolution of a composite system under oblique, positive interaction (model A, Table 1). The remnant expands onto a shelf-like density gradient with the most-dense region located in the upper left. The pulsar moves directly toward the right of the frame with a significant kick of $350\ \text{km s}^{-1}$. (a) The reverse shock interaction stage, where the expanding PWN (blue) begins an asymmetric collision with the reverse shock. (b) The PWN crushing phase, where the reverse shock rapidly compresses the PWN from all directions. (c) The slow expansion stage, where the pulsar wind slowly re-inflates the relic PWN. Tick mark spacing represents a distance of 5 pc, pulsar location is marked by a black cross.		\label{modelA}}
\end{figure*}

Following methodology used in vSDK, the pulsar kick is added by initializing the system in the pulsar's frame of rest. Here, both the pulsar and the PWN are assumed stationary, while the SNR blastwave is initialized with a velocity $-\mathbf{v}_\text{kick}$. Such a frame allows the inner grid boundary to be used as a spherically symmetric injection point for the pulsar's spin-down power.

As a final note, VH1 is purely hydrodynamical and does not assume a pulsar magnetic field. While we do expect a strong magnetic field to suppress dynamical instability at short wavelengths, the formation of longer-wavelength instability should be largely unaffected \citep[see][]{2004A&ABucciantini}, leading us to believe that the large-scale structure from which we draw our conclusions would not be impacted.

\section{Results}

Overall, evolutionary stages in the asymmetric model agree with the results presented in both BCF and vSDK. PWN morphology is heavily dependent on parameter space, so we have selected five models to highlight various features which can develop (available in Table \ref{parameters}). Specifically, the level of complexity observed in PWN morphology is dictated by the level of interaction between the BCF and the vSDK models, which we will characterize as follows: a positive (negative) interaction is one which sees the pulsar traveling toward (away from) the high-density region of the ISM, and an aligned (oblique) interaction is one in which the pulsar is moving into (tangential to) the gradient. Our use of these terms is with respect to kick angle $\psi_\text{kick}$, defined as the angle between $\bold{v}_\text{psr}$ and the density gradient $\nabla\rho$.

\subsection{General Evolution (Models A \& B)}

The PWN system presents three distinct stages of morphology: supersonic expansion, reverse shock interaction / PWN crushing, and slow expansion. The first stage is expectedly similar to the BCF and the vSDK models, where the unshocked ejecta interior is unaffected by warping of the outer blastwave. Significant changes to morphology begin to arise during the second stage, where the reverse shock begins an asymmetric crushing of the PWN. At this time, unique structure begins to form within the remnant and will persist through the slow expansion stage.

The standard evolutionary process is illustrated in Figure \ref{modelA}, where the supersonic expansion stage has been omitted. The figure follows model A, Table 1, where parameters for the composite SNR were chosen to produce an oblique, negative interaction between the BCF and vSDK models. Prior to Figure \ref{modelA}(a), the remnant has expanded rapidly from the center of the ISM density gradient, onto the density shelf. This has caused both the outer shell (red/yellow) and the reverse shock to converge to the dual hemispherical geometry seen. As the outer shell warped, the PWN (blue) was rapidly expanding within the unshocked interior until a collision with the reverse shock occurred.

The beginning of the reverse shock interaction stage is captured in Figure \ref{modelA}(a), where the pulsar's $350\ \text{km s}^{-1}$ kick has already translated the PWN enough to create a departure from symmetry. The PWN will continue to expand radially until the entire PWN shock has met with the reverse shock (Fig. \ref{modelA}(b)). Once this occurs, the PWN will be crushed as high pressure continues to drive the shocked ejecta inward. Asymmetry within the remnant will amplify as ejecta and PWN material begin to mix during the crushing, until the compressed PWN builds enough pressure to halt the reverse shock.

Figure \ref{modelA}(c) displays the end of this crushing, where a wide, tail-like structure has formed around the pulsar trailing toward the bottom left of the object. Such an observational feature may lead one to suspect that the pulsar's trajectory contains a strong upward component despite a kick directly toward the \textit{right} of the image. From this point, the pulsar will continue to move toward the right of the image as the PWN slowly re-inflates during the slow expansion stage. It is unlikely the pulsar will separate completely from the nebula due to the negative interaction between the gradient and the kick.

\begin{figure}[b]
   \centering
   \includegraphics[width=3in]
      {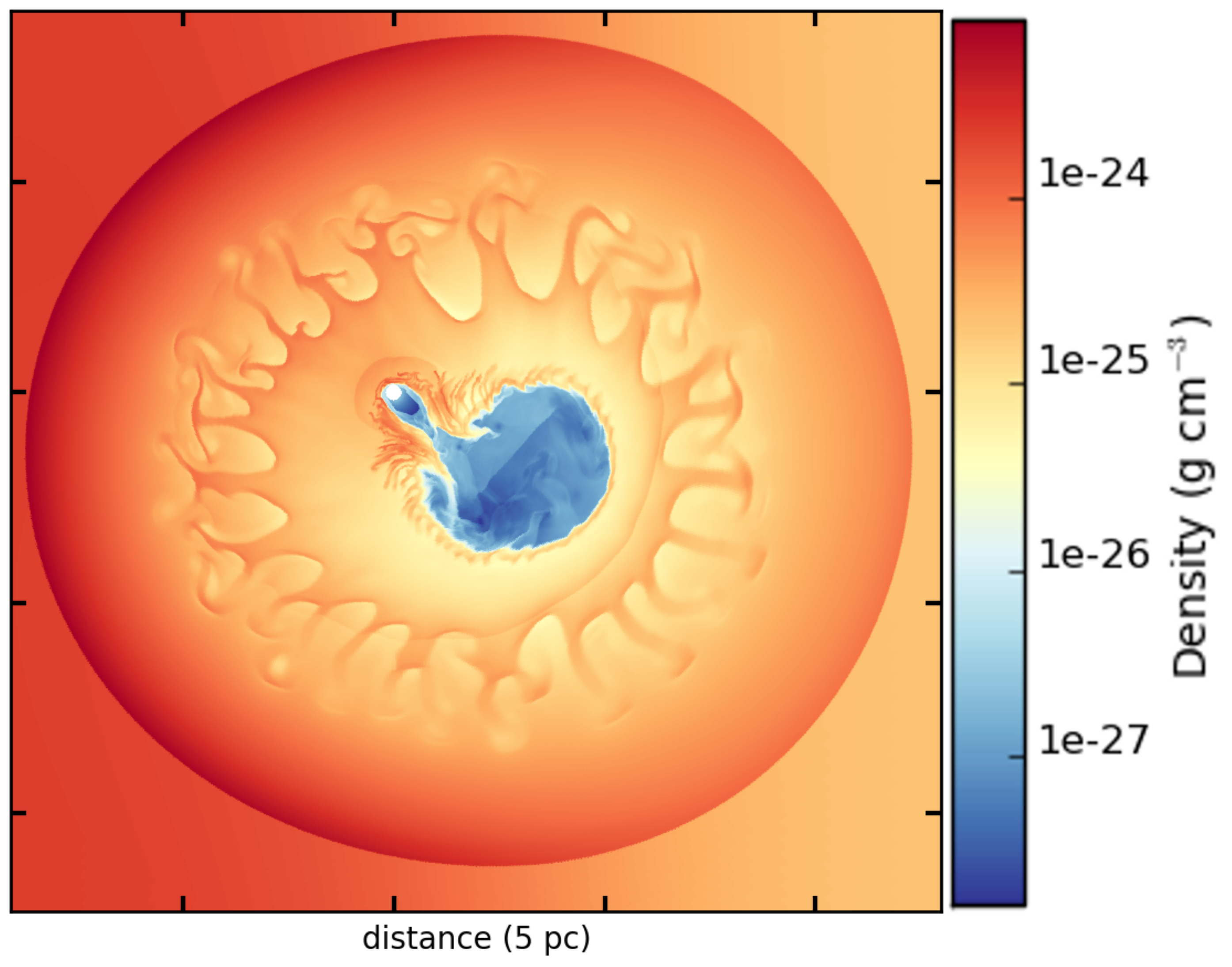}
   \caption
      {Separation of a pulsar from its PWN. Positive interaction created by the pulsar's motion (upper left direction) into the dense region of the ISM causes the reverse shock to sweep PWN material away from the pulsar as it passes, separating the pulsar from its PWN and leaving a thin tail leading to the crushed PWN.	\label{modelB}}
\end{figure}

\begin{figure*}[t]
   \centering
   \includegraphics[width=0.8\textwidth]
      {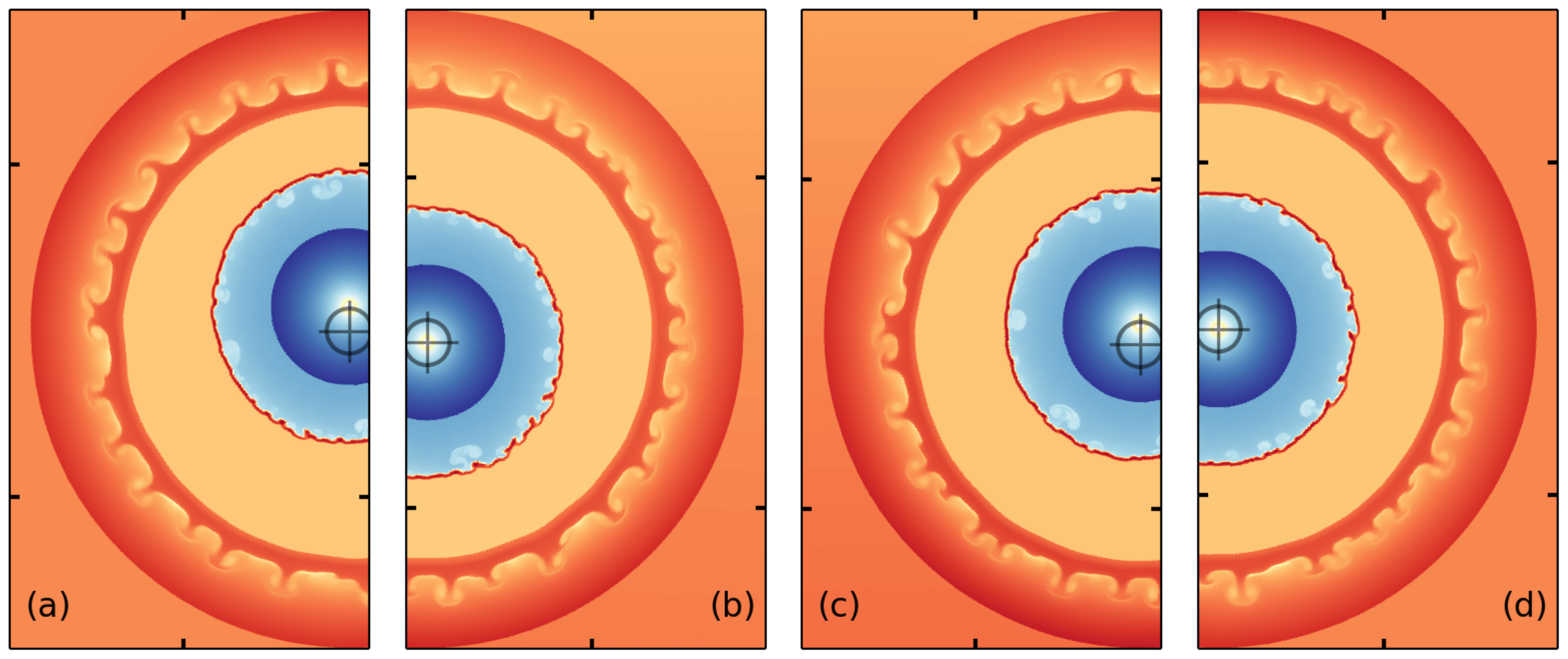}
   \includegraphics[width=0.8\textwidth]
      {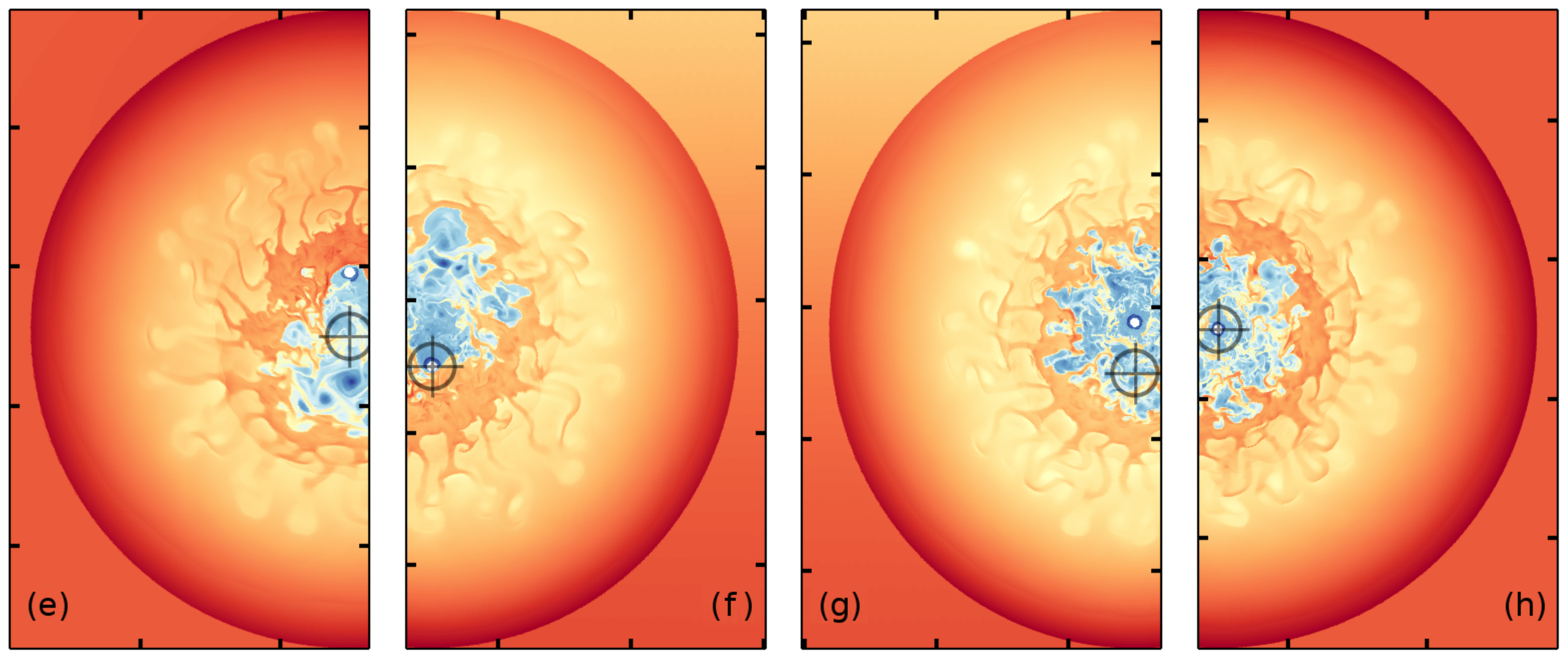}
   \caption
      {Comparison of negative, aligned interaction to parent models. 
      \textit{Top row:}
       Early stage interaction (approximately 1.2 kyr). (a) The vSDK model, with pulsar maintaining a $300\ \text{km s}^{-1}$ kick and blastwave expanding into a zero gradient. (b) The BCF model, with pulsar given zero kick and blastwave expanding into a linear density gradient. (c) Current model, with pulsar maintaining a $300\ \text{km s}^{-1}$ kick and blastwave expanding into a linear gradient. (d) Spherical-case comparison, with both zero kick and zero gradient.
      \textit{Bottom row:}
       Late stage interaction (slow expansion stage, at approximately 7.5 kyr); frames presented in same order as the above row. 
       For the given parameters, PWN morphology in the parent models (e,f) hold remarkable symmetry throughout each evolutionary stage. The model combination creates a visual replica of spherical-case evolution, where the upward moving pulsar (represented by solid white dot) remains centered within a spherical PWN, which itself remains centered with a spherical blastwave.
       Tick mark spacing represents a distance of 5 pc; origin (represented by crosshairs) has been shifted for the two middle frames in each row.
	\label{modelC} }
\end{figure*}

\begin{figure*}
   \centering
   \includegraphics[width=0.8\textwidth]
      {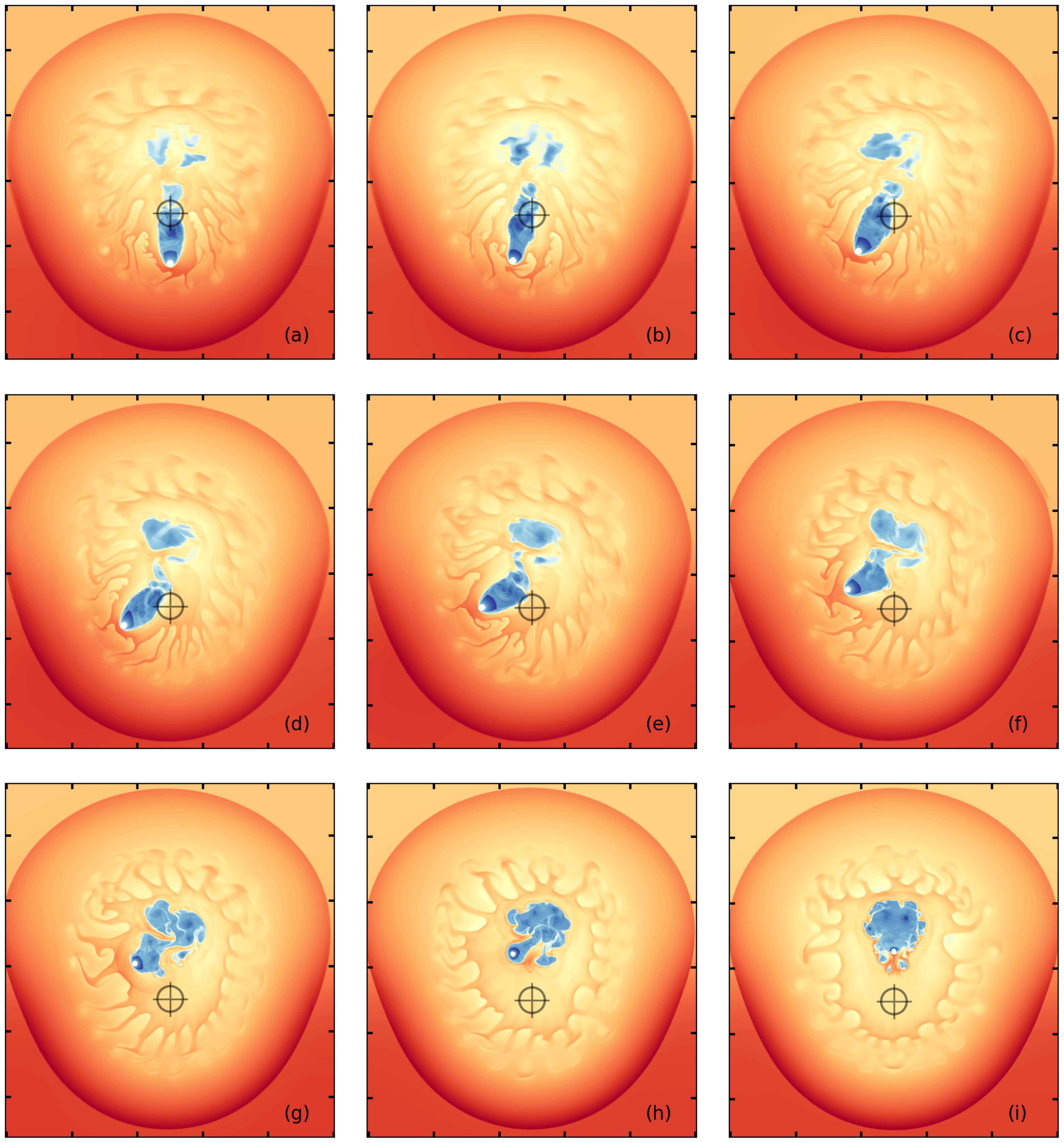}
   \caption
      {PWN morphology for model D, with varying kick trajectories. $\psi_\text{kick}$ increases in increments of $\pi / 8$ radians in between frames, from $\psi_\text{kick} = 0$ in (a) to $\psi_\text{kick} = \pi$ in (i); all other parameters are held constant. The pulsar (white circle) is located at the head of the PWN tail (blue region) in each frame. The explosion center has been marked by crosshairs to help visualize kick trajectory. Spacing between tickmarks represents a distance of 5 pc.
	\label{modelD} }
\end{figure*}

To provide contrast, model B employs an aligned, positive interaction between the BCF and vSDK models. The characteristic length of the ISM density gradient $H$ is chosen such that the blastwave of model B remains in the linear region of the gradient until the system evolves past the PWN crushing stage. Within this region, the outer shell warps over a significantly longer timescale, which allows both the blastwave and the reverse shock to retain near-spherical geometry. The unshocked interior of the SNR is unaffected by this minor warping during the supersonic expansion stage, leading the system to behave as a direct linear combination of the two models.

A positive interaction guarantees that the first point of collision between the PWN and the reverse shock falls in line with the pulsar's direction of travel. This collision marks the beginning of asymmetric evolution within the model. (Although not immediately apparent in various figures, a large portion of the linearity seen in the supersonic expansion phase persists through the system's chaotic evolution and is discussed in detail in the following section.)

The most notable result of this parameter combination can be seen in Figure \ref{modelB}, where a defined tail has formed as the reverse shock crushes the PWN. Similar behavior is observed in the two underlying models: The BCF model has a tendency to sweep PWN material toward the side of the SNR located in the low-density portion of the ISM, leaving a stationary pulsar located either near the edge or just outside of the relic PWN. The vSDK model observes a similar placement of the pulsar as the kick moves the pulsar toward one edge of the spherical PWN; resulting in the formation of a wide, symmetric tail as the reverse shock is swept aside by the pulsar's strong termination shock.

These effects are additive for the positively interacting combined model, and the formation of a defined tail is always probable. Once it has formed, the tail will develop a slight deflection away from the path of the pulsar's proper motion as well as a slight curvature toward the PWN. For a particularly violent crushing of the PWN, the same mechanism creating this curvature may instead sever the newly formed tail, effectively separating the pulsar from its PWN. This separation will occur in model B shortly after the time-step from Figure \ref{modelB}.

The method of subsonic expansion during the final stage will depend on the formation of a tail during PWN crushing. The pulsar in a negatively interacting system normally remains within its PWN and is not likely to see the formation of a tail. Such a system allows for direct inflation of the relic PWN by the pulsar. Systems with oblique positive interaction are likely to see a tail formed without separation, leading to an indirect inflation of the relic PWN as new gas travels the tail and into the PWN. Aligned positive interactions are most likely to experience a full separation between the pulsar and the PWN, resulting in the formation of a small, secondary PWN which inflates surrounding the pulsar.

\subsection{Negative Interaction (Model C)}

Parameters for model C have been chosen to demonstrate both the behavior of the negative interaction model and the predictability of a linear ISM gradient. The resulting behavior is displayed rather suggestively in Figure \ref{modelC} as each of the parent models are placed alongside both the combined model and a spherical-case simulation employing the same parameters. Figure \ref{modelC}(a) contains the vSDK model 1.2 kyr after the supernova event; given an upward kick of $300\ \text{km s}^{-1}$. Figure \ref{modelC}(b) shows the BCF model with gradient strength chosen to produce an effective remnant velocity of $\sim 300\ \text{km s}^{-1}$, moving the geometric center upward and away from the stationary pulsar. We note in each frame, the spherical PWN remains centered around the pulsar rather than centered within the remnant shell. Crosshairs are placed in each frame to visibly mark the explosion center.

Figure \ref{modelC}(c) displays the combined system using both the $300\ \text{km s}^{-1}$ kick and the linear gradient. The two models are seen to combine very linearly during the supersonic expansion stage; the upward motion of the shell center follows directly with the upward motion of the kick. (We note that net velocity is slightly biased toward the kick direction during the first 300 yr of the simulation, as the initially spherical blastwave requires time to warp and reach an equilibrium state.) The shell is also seen to remain approximately spherical while the system is contained in the linear region of the ISM gradient. For visual comparison, figure \ref{modelC}(d) displays the spherical case, where both kick and gradient have been removed.

Late-stage evolution for Model C is shown in Figure \ref{modelC}(e-h), approximately 7500 yr after the supernova event. The system has entered the subsonic expansion stage, and the two parent models are nearly identical mirror images of each other. Figure \ref{modelC}(e) has the relic PWN positioned near the center of the remnant shell with the pulsar located at the upper edge due to the kick. In contrast, Figure \ref{modelC}(f) sees the stationary pulsar positioned near the lower edge as a result of the PWN shifting upward toward the remnant center during the reverse shock interaction stage.

The significance of this model is seen in Figure \ref{modelC}(g), where a combination of the kick motion and the effective remnant velocity places the pulsar at the center of the PWN, which is itself positioned at the center of the still-spherical remnant. For comparison, Figure \ref{modelC}(h) demonstrates how closely negative interaction can resemble the spherical model; implying visual symmetry does not necessarily limit the age of an object nor does it exclude the possibility of a significantly fast kick. In general, negative interaction has a tendency to keep the pulsar located within a PWN during the reverse shock interaction, limiting the likelihood of a separation event.

\begin{figure}[b]
   \centering
   \includegraphics[width=3in]
      {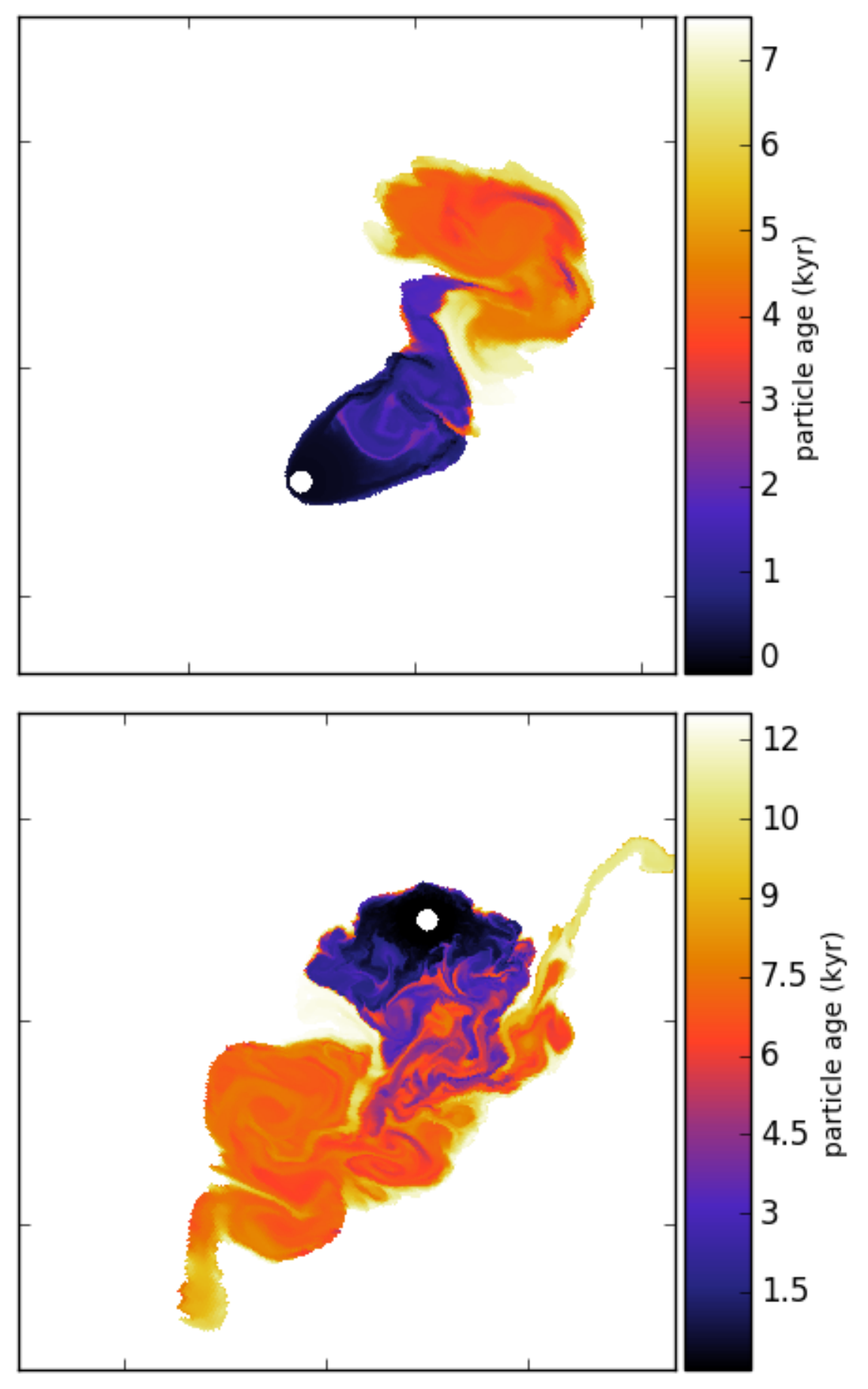}
   \caption
      {Anisotropic behavior of particle age for various models. (a) Model D at 7.5 kyr; freshly injected gas (black) located near the pulsar (white circle) can be seen progressing through the wide PWN tail, with the oldest gas (orange/yellow) located in the body of the relic PWN (PWN morphology can be seen in figure \ref{modelD}(g)). (b) Model A at 12.5 kyr; freshly injected gas can again be seen near the PWN, with the age variable increasing alongside distance from the pulsar (PWN morphology available in figure \ref{modelA}(c)). In each case, a strong directional dependence has developed. 	\label{gasage}}
\end{figure}

\begin{figure*}[t]
   \centering
   \includegraphics[width=\textwidth]
      {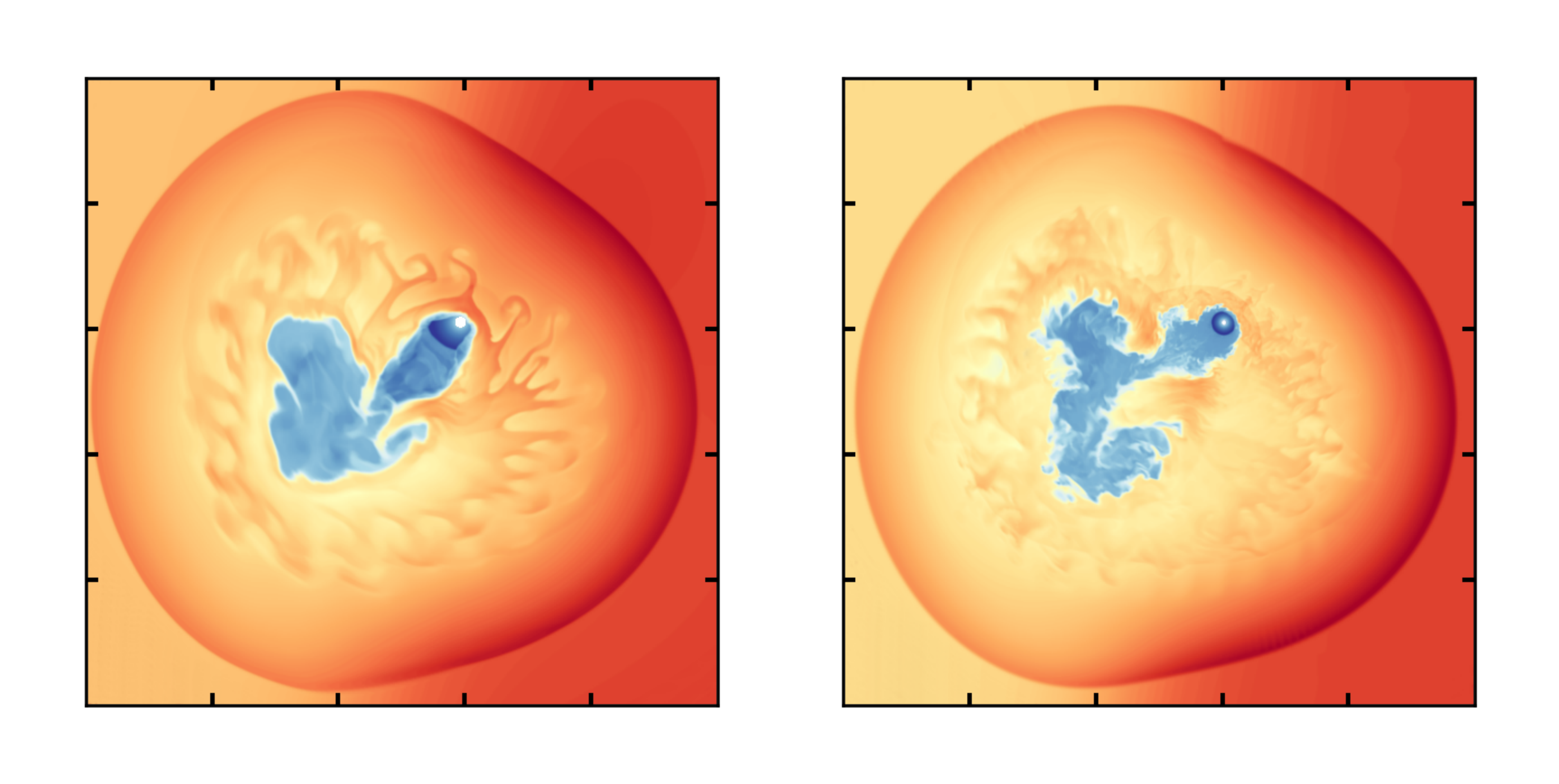}
   \caption
      {Comparison of morphology for the 2D model [\textit{left}] and the 3D model [\textit{right}] for SNR G327.1-1.1 (Table 1, model E), evolved to 17.4 kyr. Simulation for the 2D model is performed on a cylindrical grid with an angular resolution of 0.625 degrees. Simulation for the 3D model is performed on a spherical Yin-Yang grid with an angular resolution of 0.6 degrees; image is produced by taking an equatorial slice of the grid. 	\label{Ecomparison}}
\end{figure*}

\subsection{Directional Dependence (Model D)}

A system with positive, aligned interaction is likely to form a PWN `tail' during the passage of the reverse shock. This tail manifests as a thin trail of fresh PWN material linking the active source to a displaced relic nebula, and its formation occurs as incoming shocked SNR ejecta is forced around the head of the PWN by the pulsar's termination shock while the bulk of the reverse shock continues to crush the PWN.

Figure \ref{modelD} displays a system given a large kick of $|\mathbf{v}_\text{kick}| = 500\ \text{km s}^{-1}$, where $\psi_\text{kick}$ has been varied in increments of $\pi / 8$ radians; all other parameters held constant. Crosshairs again mark the pulsar's origin and may be used to visualize the direction of travel. The pulsar in Figure \ref{modelD}(a) travels downward, directly into the gradient with a kick angle of $\psi_\text{kick} = 0$, and a large tail is seen to develop as the pulsar makes its way through the reverse shock.

With an increase in the value of $\psi_\text{kick}$, the tail develops a visible deflection away from the point of origin. Once $\psi_\text{kick}$ exceeds $\pi/2$ radians, the tail shows significant deflection relative to the object's motion. Such a result reveals immediately that the observed direction of a tail produced in a system evolving within a large density gradient is a poor indicator of the pulsar's proper motion. Behavior of the deflection also exposes a significant mechanism present within these models:

Seen in each of panels (a-i) in Figure \ref{modelD}, the relic PWN always tends to reach an equilibrium position on the side of the remnant located within the low-density region of the ISM. This behavior is discussed briefly in BCF, where the movement is attributed to the development of a pressure gradient behind the reverse shock. In each instance, the PWN tail leads to the side of the relic PWN closest to the high-density region of the ISM (the bottom edge of the relic in this example). The mechanism behind this behavior is discussed more thoroughly in Section \ref{ssec:PWNtail}.

\subsection{Gas Age}	\label{ssec:gasage}

Tracking the motion of PWN gas is a useful tool when investigating morphology. The addition of a particle-age variable simplifies this processes by generating a grid-zone map which depicts time since gas has passed through the termination shock. This variable behaves exactly as one would expect, with the majority of aged particles appearing within the PWN, furthest from the emission source (see Fig. \ref{gasage}(a)). For the case of a PWN which has formed a tail, the variable is able to demonstrate the path particles take as they travel from the emission source, through the tail, and into the relic PWN. Figure \ref{gasage}(b) demonstrates this, where freshly injected particles (black) are seen at the head of the PWN while aged particles (yellow) have moved down the tail into the relic where the gas begins to mix. Any system which evolves asymmetrically will develop an anisotropic particle age map once the reverse shock sweeps PWN material away from the geometrically off-centered injection point.

\subsection{Three-Dimensional Evolution of cSNR G327-1.1-1 (Model E)}	\label{ssec:modelE}

We have selected the composite remnant G327.1-1.1 to test the validity of both the 2D and 3D models. The 2D model has been applied to G327.1-1.1 in an earlier paper by \cite{2015ApJTemim}, where the authors were able to produce PWN structure similar to the observed radio relic PWN; allowing the authors to infer several physical constraints on the system. Notably, they were able to approximate the age of the remnant (an estimated 17.4 kyr) and a likely high-to-low ISM density contrast ratio and orientation. Additionally, \cite{2015ApJTemim} were able to successfully apply the particle age-tracking variable mentioned in section \ref{ssec:gasage} to explain the morphology seen in both X-ray and radio observations of G327.1-1.1. In the paper, simulation was able to produce an object very similar in morphology to the observed remnant: The model develops a trail consisting of freshly injected particles, which agrees nicely with observed X-ray emission. Additionally, the separated relic PWN consists almost entirely of aged particles, implying they have incurred significant synchrotron losses and are expected to produce a strong radio emission.

Evolution of the 3D model of G327.1-1.1 presented in this text remains largely consistent with these results, and more generally, properties of the 3D model remain consistent with the generic 2D models A-D presented here (particularly in regards to overall evolution, where the same three stages are visited and all timescales remain unaffected).

\begin{figure}[b]
   \centering
   \includegraphics[width=3in]
      {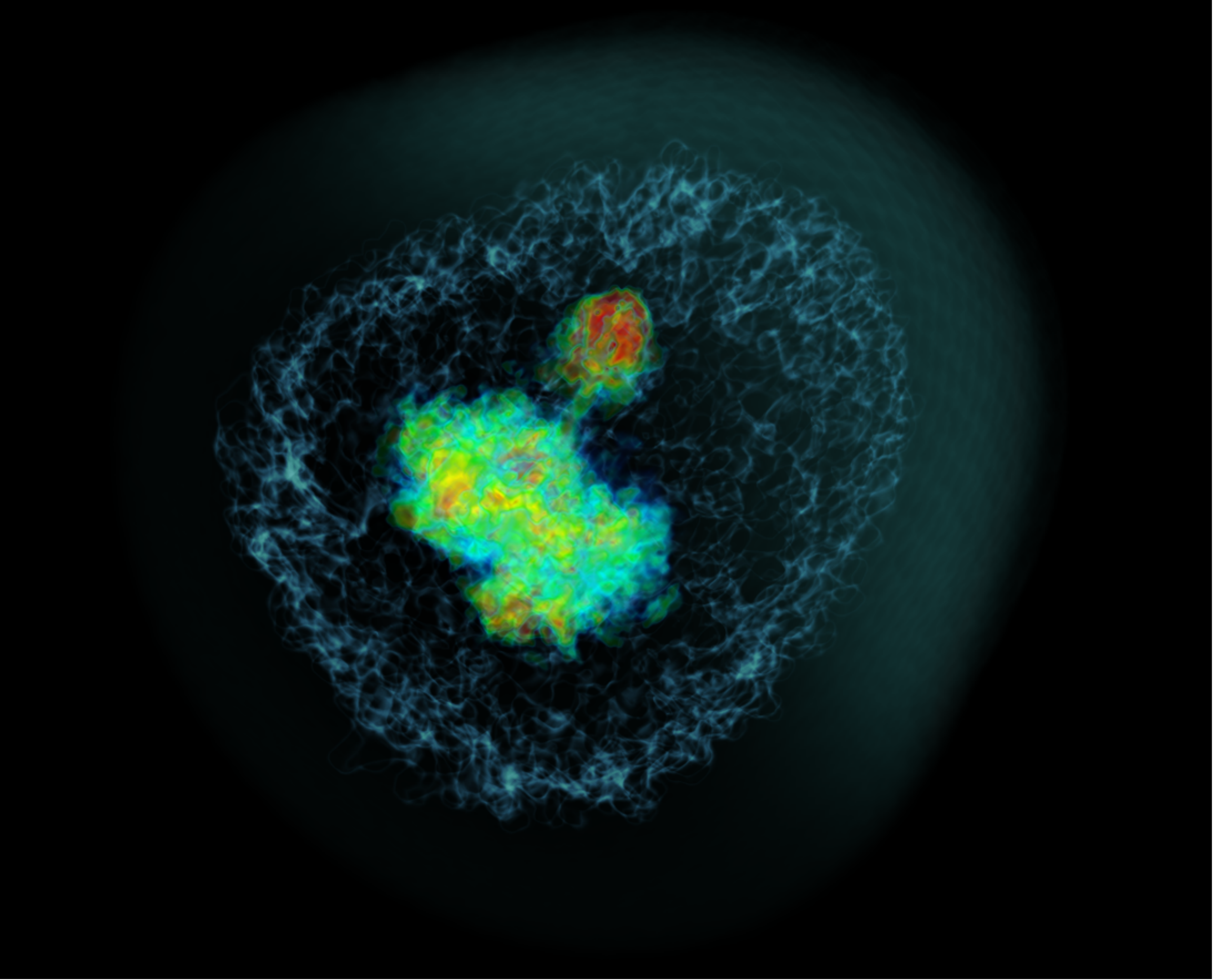}
   \caption
      {3D volume rendering of model E (Table 1) evolved to 17.4 kyr, with pulsar traveling in an upward direction. The faint outer shell depicts the blastwave, while the inner instability (light blue) shows the contact discontinuity between shocked ejecta and ISM within the remnant. The remnant's interior hosts the PWN, which has been colored by gas composition (where a composition of 100\% PWN gas by mass has been colored red, 50\% PWN to 50\% shocked ejecta by mass has been colored green, and 10\% PWN to 90\% shocked ejecta by mass has been colored blue). An interactive version of this model is available at \url{http://astro.physics.ncsu.edu/~cekolb/pwnsnr/x3d/vol.xhtml}, which will allow the reader to rotate the object and explore the various shapes this PWN can take when viewed from different angles.
	\label{modelE}}
\end{figure}

\begin{figure}[b]
   \centering
   \includegraphics[width=3in]
      {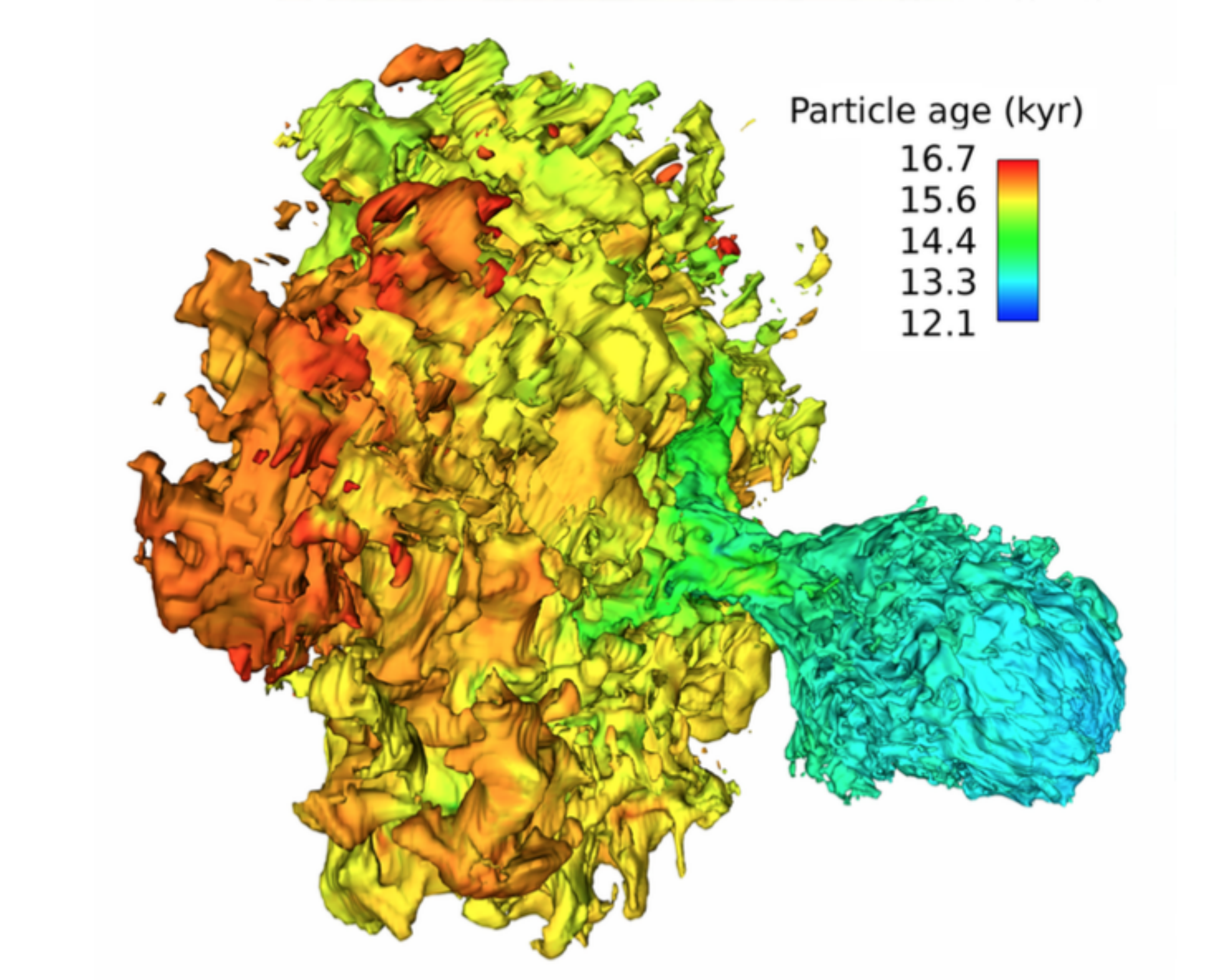}
   \caption
      {A rotated view of the PWN in Figure \ref{modelE}, colored according to average particle age (pulsar motion toward the bottom-right). The solid isosurface represents zones containing $\> 20\%$ PWN material by mass. Young PWN gas can be seen flowing from the head of the object (blue) through the tail (green), and into the relic PWN (yellow/orange). Not visible is a blue sphere of young gas within the head of the nebula, containing the pulsar with termination shock.	\label{pwn_age}}
\end{figure}

The two models are compared visually in Figure \ref{Ecomparison}, where an equatorial slice of the 3D model is shown beside the 2D model at the age we believe most appropriate for G327.1-1.1 (17.4 kyr, presented in \cite{2015ApJTemim}). The most apparent similarities between the two models are the geometry of the outer blastwave and the contact discontinuity. Symmetry of the density gradient allows the 3D shell geometry to be extrapolated from the 2D simulation by a simple rotation around the gradient's $z$-axis. The contact discontinuity resembles the geometry of the blastwave as seen in \citep{2013MNRASWarren}, with more complexity resulting from fluid mixing in the additional dimension. This added complexity is visible in Figure \ref{modelE}, where a 3D volume rendering of the PWN, contact discontinuity, and the outer blastwave may be seen. Additionally, behavior of the particle age variable remains consistent with the 2D model, with the oldest particles collecting in the relic PWN. Interestingly, the additional dimension leads the majority of aged particles to congregate on the outer surface of the PWN, while freshly injected particles flow through the interior of the PWN tail into the central region of the relic. A closer look at the age of particles in the PWN is available in Figure \ref{pwn_age}, where a surface representing $\> 20\%$ PWN gas by mass has been colored according to particle age.

Three major differences between the 2D and 3D models arise during the reverse shock interaction stage. Most noticeable is the direction of the tail, which is seen in Figure \ref{Ecomparison} to point drastically more toward the left in the 3D figure than in its 2D counterpart. (This is a direct result of the dimensionality and is discussed further in Section \ref{ssec:PWNtail}.) Secondly, the termination shock surrounding the pulsar is significantly more spherical in 3D than in the various 2D models presented here as well as in vSDK and similar papers. This is also believed to result from the dimensionality, which seems to restrict the formation of stable triple-points seen in 2D, opposite the direction of the incoming reverse shock. Lastly, the PWN crush resulting from the reverse shock interaction appears to be significantly more violent; again explained by the dimensionality as ejecta are more easily able to mix with the PWN rather than compressing it. It is important to note that apparent morphology for the 3D system may be drastically altered by a simple change in viewing perspective.

\section{PWN Morphology}

\begin{figure*}[t]
   \centering
   \includegraphics[width=\textwidth]
      {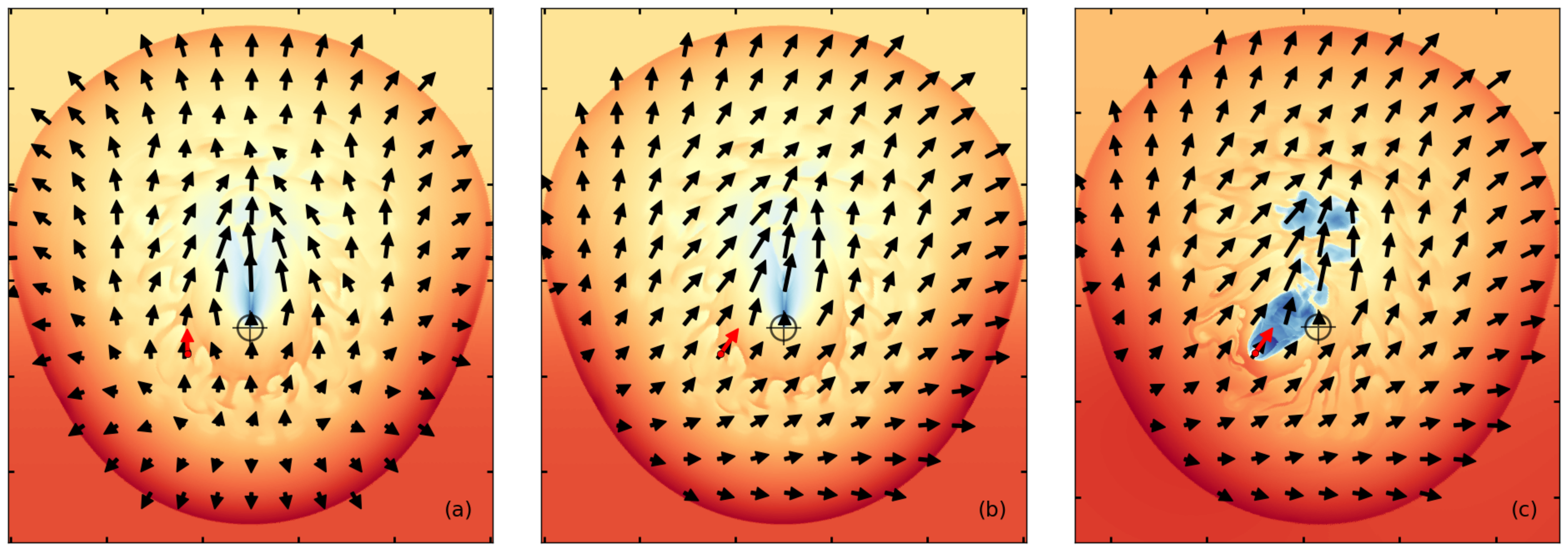}
   \caption
      {Simulation method used to predict PWN tail direction. (a) Ejecta velocity flow field for a lone SNR in a stationary reference frame (simulation based on parameters in Table 1, model D); red arrow marks velocity vector at the prediction point $\mathbf{v}_\text{psr} t$. (b) The same flow field, transformed to a reference frame moving at velocity $\mathbf{v}_\text{psr}$ (with $\psi_\text{kick} = 3\pi / 8$ radians). (c) The same transformed flow field shown in (b), overlaid onto a separate, composite simulation of model D. A strong agreement between the ejecta velocity flow field at the prediction point (red arrow) and the actual direction of the PWN tail can be seen.	\label{vfield}}
\end{figure*}

The development of PWN asymmetry is dependent on three main components of the composite system: the evolutionary timescale $t_\text{ST}$ \citep[see characteristic timescale,][]{1995PyRepMcKee}, the pulsar spin-down power, and the strength of interaction between the pulsar kick velocity and the external density gradient. Evolutionary timescale proves to be the weakest component for any system using realistic parameters, as transitions between the supersonic expansion and the reverse shock interaction stages depends heavily on the speed and direction of the pulsar kick. Additionally, $t_\text{ST}$ primarily serves to determine the contribution that other parameters will have on morphology (e.g., a large $\tau / t_\text{ST}$ ratio will place more emphasis on spin-down power, leading to a larger relic and a thicker tail). For a linear gradient with an average ambient density of $\rho_0$, the evolutionary timescale closely follows the timescale of a remnant evolving in a uniform ISM of density $\rho_0$ (see Fig. \ref{modelC} for results of a system with an optimally chosen $\mathbf{v}_\text{psr}$). As the gradient becomes more shelf-like ($\nabla\rho \to \infty$), the remnant develops a bihemispherical geometry which will add large variation to this timescale.

The pulsar's spin-down power can have a moderate effect on asymmetry. The primary contribution by this term is through the total energy available to the PWN during the crush phase (found by integrating $\dot E(t)$ from equation \eqref{Edot}), which largely determines the final volume of the crushed PWN (BCF). This, in turn, determines the potential for tail formation, as a smaller nebula is more likely to separate from the pulsar. Such an effect can be seen by comparing models A and D, Table 1. Model A (large $\tau/t_\text{ST}$ ratio) develops a relatively large PWN from which the pulsar is unable to exit, while Model D (small $\tau/t_\text{ST}$ ratio) develops a much smaller PWN which easily allows the pulsar to separate. (See Fig. \ref{modelD} for reference.)

The strongest impact on PWN asymmetry develops from the interaction strength between the BCF and vSDK models. Significant variation in morphology can arise by altering this interaction while holding all other parameters equal (e.g., Fig. \ref{modelD}). A kick running oblique to the gradient (Fig. \ref{modelD}d-f) will generally maximize this effect, while a kick aligned with the gradient will result in a relatively straight tail leading to a spherical relic PWN.

\subsection{Formation and Deflection of the PWN Tail}  \label{ssec:PWNtail}

The formation of a tail is likely for a system with a degree of positive interaction between the BCF and vSDK models, but may occur in any system which produces a large displacement in PWN location. The eight simulations of Model D which produced a tail (Fig. \ref{modelD}a-h) reveal a non-linear relationship between the direction of the tail and the true motion of the pulsar; this deflection angle of the tail $\theta_D$, measured from the true motion path, is shown to increase as a function of the object's trajectory angle $\psi_\text{kick}$.

Both the formation of this tail and its deflection are a direct result of the interaction between the density gradient and the pulsar's velocity. This property is readily explained by the significantly limited spin-down power driving the PWN during the crush phase (where $t > \tau$ and $L_\text{p}(t) \to 0$); a significantly displaced pulsar which interacts with the reverse shock is only able to disrupt a small portion of the incoming shock. As a result, the bulk of the shocked ejecta continues to flow uninterrupted toward the explosion point, and a PWN tail is swept out as a small amount of material is forced around the pulsar/termination shock.

Analysis of model D reveals the deflection angle to hold a large degree of predictability, based on the blastwave geometry developed by the density gradient. Figure \ref{vfield} demonstrates this in a three-step process:

\begin{enumerate}
\itemsep0em
\parskip0em

\item An SNR is evolved in a stationary reference frame, with pulsar + PWN excluded from the simulation; all parameters from the model of interest are held constant. Fig. \ref{vfield}(a) shows the velocity vector plot for such a simulation, at a time $t = 12.5$ kyr.

\item The velocity plot from the previous step is transformed to a frame moving with velocity $\mathbf{v}_\text{psr}$ (Fig. \ref{vfield}b). At this point, the ejecta velocity local to the position $(\mathbf{v}_\text{psr}\ t)$ can be used to predict tail direction; shown in the figure with a red arrow. We note that this prediction method will develop error for points of interest located near the remnant's turbulent contact discontinuity. 

\item Curvature of the tail and approximate location of the relic PWN can be determined by following the vector field from the location of the pulsar. As evidence, Figure \ref{vfield}(c) shows the same vector field from the previous step, overlaid onto one of the full model D simulations. A direct overlay such as this will also produce some inaccuracy when predicting tail curvature, as this is a function of the `history' of ejecta flow. 
\end{enumerate}

\begin{figure}[b]
   \centering
   \includegraphics[width=3in]
      {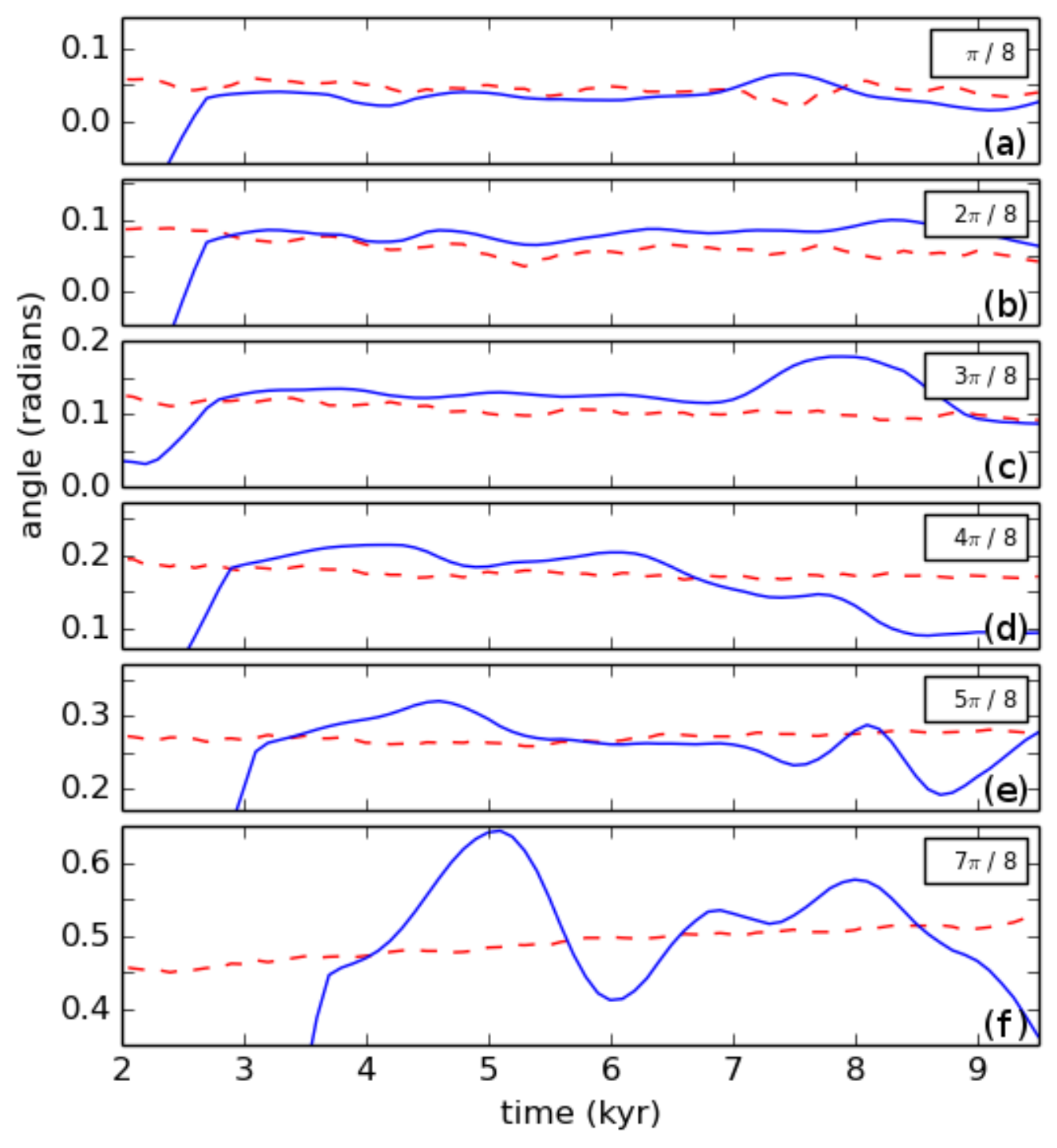}
   \caption
      {Tail deflection angle (blue) and angle of local ejecta current (red) for various kick trajectories of model D. Systems with strong positive interaction (a-c) show little deviation from ambient ejecta flow, while systems with negative interaction (d-f) reveal large oscillations about this radial flow.	\label{tailangle}}
\end{figure}

\begin{figure*}[t]
   \centering
   \includegraphics[width=\textwidth]
      {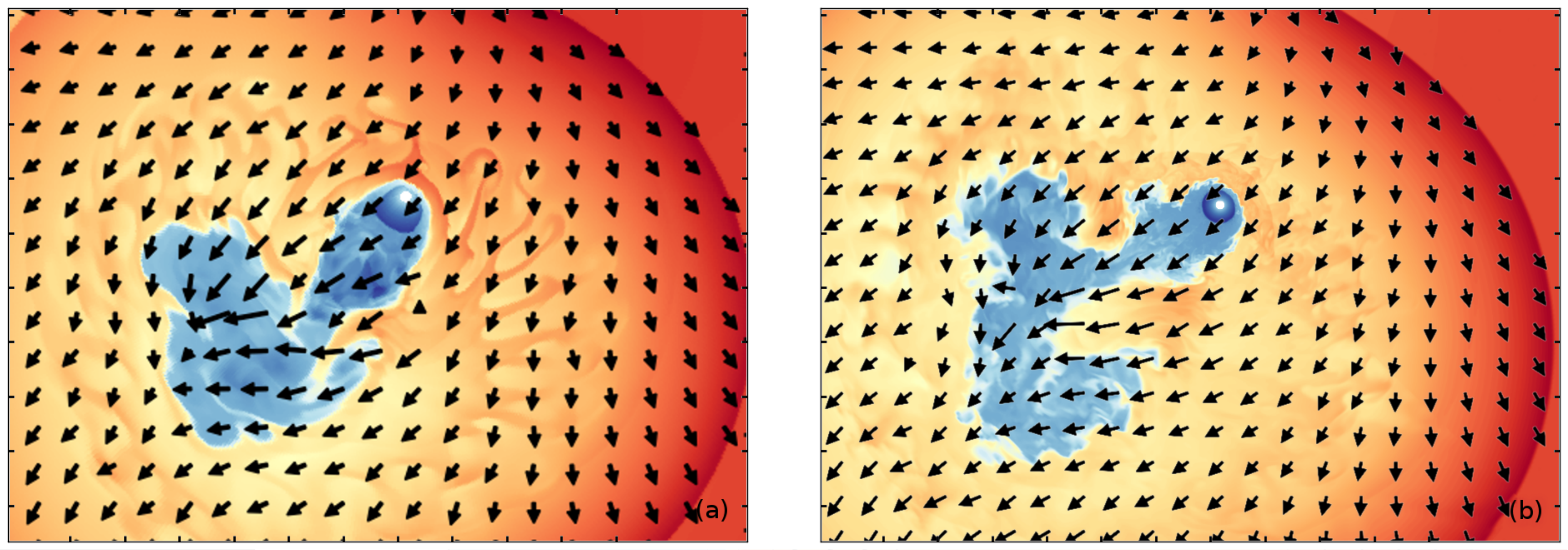}
   \caption
      {The same predictive method used in Figure \ref{vfield} applied to the system G327.1-1.1 (Table 1, model E). (a) The ejecta velocity flow field resulting from the 2D model, overlaid on to the 2D density plot. (b) The ejecta velocity flow field resulting from the 3D model, overlaid onto a density plot taken in the equatorial plane of the 3D simulation. Frame of reference for each flow field is that of the moving pulsar.		\label{g327vfield}}
\end{figure*}

Applying this process to several models demonstrates a strong correlation between tail direction in the composite system and the local velocity vector field of shocked ejecta within an isolated SNR system. The direction of the tail is quantified by taking a particle-age weighted average of gas velocity angle and is shown in Figure \ref{tailangle} (blue line), with a cutoff placed at one-tenth the simulation age in order to avoid weighting material located inside the relic PWN. The local ejecta velocity vector (dashed red line) is then determined by taking a distance-weighted average in the isolated SNR model using the zones nearest the $(\mathbf{v}_\text{psr} t)$ position and then transforming to a reference frame moving at $\mathbf{v}_\text{psr}$; noise present is due to turbulence behind the reverse shock. (Note that this method uses velocity of the portion of the reverse shock nearest to the point of interest $\mathbf{v}_\text{psr} t$ for time-steps at which the reverse shock has not yet passed $\mathbf{v}_\text{psr} t$.)

Seen in each subplot of Figure \ref{tailangle}, the deflection angle has a sharp jump from $\theta_D \sim 0$ radians (while the pulsar is located at the center of the PWN during the supersonic expansion stage) to match the direction of the local ejecta velocity field as the reverse shock passes by the pulsar and a tail is formed. The deflection angle has an oscillatory behavior which grows stronger in amplitude as the model moves from positive to negative interaction. This growth in amplitude is explained by a sloshing motion seen in the PWN (previously noted in BCF), which will dominate tail deflection angle for systems with strong negative interaction between the kick and the ISM density gradient. This is witnessed in Figure \ref{tailangle}, where an aligned, positively interacting system (subplots a-c) holds a relatively constant deflection angle oscillating within a few degrees of the local ejecta velocity vector. For obliquely interacting systems (subplots d-f), larger oscillations about this vector occur.

Additionally, this prediction method holds true when moving to three dimensions, where it is able to explain the significant increase in tail deflection angle noted for model E in Section \ref{ssec:modelE}.  Shown in Figure \ref{g327vfield}, the extra dimensionality produces an ejecta flow field near the SNR interior with a much stronger leftward component than is seen in the 2D model. Just below the pulsar, this flow field is nearly perpendicular to the pulsar's upward motion, and a strong sweeping effect is produced which points the tail significantly further to the left.

\subsection{Effective Velocity}

Since the exact parameters of a composite SNR system may not be known, a naive approach to quantifying PWN tail deflection may be useful. The method we have developed involves consideration of a pseudo velocity term $\mathbf{u}_\text{snr}$, defined by the motion of the SNR's geometric center with respect to the explosion point. Such a motion is the direct result of the blastwave's differential propagation as the shock expands much more rapidly into the low-density region of the ISM than into the high-density region. Consequently, the shell's geometric center moves toward the low-density side of the ISM gradient and a stationary pulsar appears much closer to the high-density side of the blastwave.

\begin{figure}[b]
   \centering
   \includegraphics[width=3in]
      {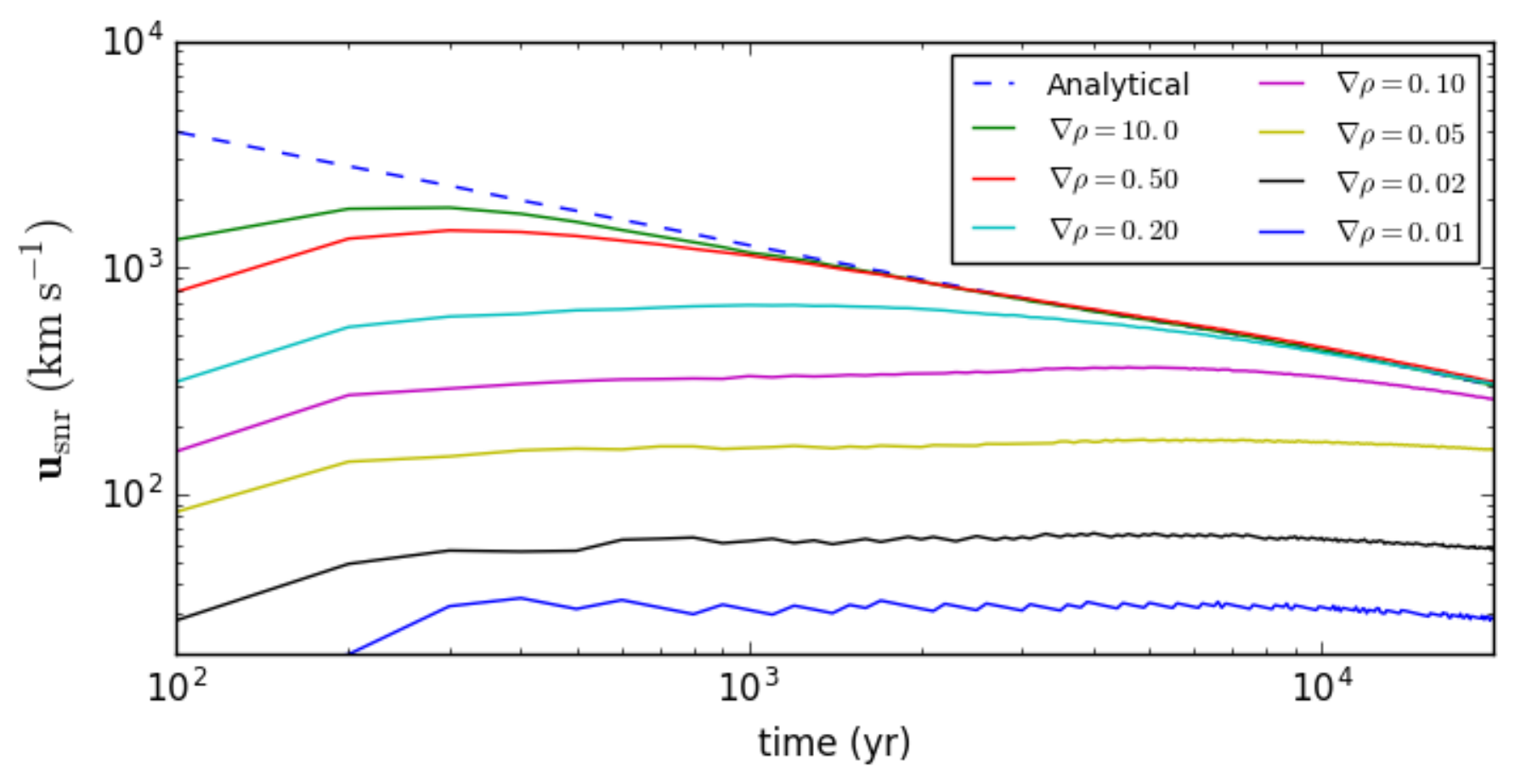}
   \caption
      {Pseudo velocity for various values of $\nabla\rho$ (units of displayed values are $\text{cm}^{-3}\ \text{pc}^{-1}$). Each requires an equilibrium timescale on the order of $\sim$ 100 yrs as the once-spherical blastwave warps to match the ISM density gradient. Once this equilibrium is reached, $\mathbf{u}_\text{snr}$ remains approximately constant until reaching the asymptote provided by the analytical solution for the shelf-like case (implying $\nabla\rho \to \infty$). Simulations presented use a density contrast of $20 \: 1$ ($x = 1.05$) with average ISM density $\rho_0 = 10^{-24}\ \text{g cm}^{-3}$. Lines appear in descending order, starting with the analytical solution (dashed) and moving down toward $\nabla\rho = 0.01\ \text{cm}^{-3}\ \text{pc}^{-1}$.		\label{usnr}}
\end{figure}

This effect combines rather linearly with the pulsar's kick velocity such that one may predict the approximate direction of the tail by taking the difference of the two. In Figure \ref{modelD}, $\mathbf{u}_\text{snr}$ points directly upward toward the low density region of the ISM, and the tail direction may be predicted in all cases by determining the effective velocity $\mathbf{v}_\text{eff} = \mathbf{v}_\text{psr} - \mathbf{u}_\text{snr}$. Thus, observation of a system with a deflected tail (providing $\mathbf{v}_\text{eff}$) and a measurable kick may help infer the presence and strength of a density gradient. (It is noted that this method is far from exact and quickly loses accuracy as the gradient becomes more shelf-like.)

Further exploration reveals a strong dependence on an analytical solution for the shelf-like gradient ($\nabla\rho\to\infty$), found by taking the difference in propagation speeds for a one-dimensional blastwave expanding on each density shelf. Figure \ref{usnr} demonstrates the asymptotic behavior provided by this solution for various values of $\nabla\rho$. For each, the initially spherical remnant requires a short convergence period of $200-300$ years, after which $\mathbf{u}_\text{snr}$ remains approximately constant while approaching this asymptote. Physically, this period of constant pseudo velocity is due to the blastwave expanding in an approximately linear density gradient, and the convergence with the analytical solution is due to the blastwave expanding well beyond this linear regime to a point where the gradient is approximately shelf-like.

The pseudo velocity term proves useful when considering evolution in a purely linear density gradient. Implications of this term are demonstrated in Figure \ref{modelC}, where the gradient has been chosen such that $\mathbf{u}_\text{snr}$ matches $\mathbf{v}_\text{psr}$. The model evolved within a gradient (Fig. \ref{modelC}g) provides a near-exact match to the uniform density solution (Fig. \ref{modelC}h) due to the pulsar's effective velocity of approximately $\mathbf{v}_\text{eff} = 0\ \text{km s}^{-1}$. For a composite system evolving in a linear gradient, evolution of the PWN may be inferred through this effective velocity by treating it as the initial kick in a uniform density system and applying methods presented in vSDK.

\section{Discussion and Conclusions}

We have considered the case of a PWN interacting with a SNR, for which the SNR blastwave is allowed to expand into a density gradient and the pulsar is given a high kick velocity at an arbitrary direction with respect to this gradient. We have accounted for energy lost to the PWN through synchrotron emission by applying a crude inverse-decay term to the pulsar spin-down luminosity. We note such a treatment of synchrotron loss is intended only to adjust energy available to the PWN during late-stage evolution and may slightly alter the timescale of the transition between the supersonic expansion phase and the reverse shock interaction phase.

Our investigation has focused around two ISM gradient geometries; linear (defined by the characteristic length scale $H > R_\text{SN}$) and shelf-like (with $H \ll R_\text{SN}$), producing a blastwave with approximate spherical and dual-hemispherical geometry, respectively. We have also considered two models of interaction between this gradient and the pulsar kick: positive (pulsar headed toward the high-density region of the ISM) and negative (pulsar headed away from the high-density region). By exploring these parameters through simulation, we have shown that a composite supernova remnant undergoing asymmetric evolution will experience the same three phases noted in \cite{2001ApJBlondin} and \cite{2004A&AVDS}: free expansion, reverse shock interaction, and subsonic expansion. While the free expansion phase remains directly consistent with BCF and vSDK, interaction between the pulsar kick and then ISM density gradient may lead to the formation of a PWN tail and a potential pulsar separation event; resulting in several possible methods of inflation during the subsonic expansion phase.

We have demonstrated that for a sufficiently strong positive interaction between the pulsar kick and the ISM gradient, the PWN tail is formed as the reverse shock sweeps PWN material away from the pulsar and leaves a trail of PWN material linking the pulsar to the relic PWN. Separation between the pulsar and the PWN occurs when the reverse shock is strong enough to sever the newly-formed tail. We have further investigated the dynamics present during the formation of this tail, and we have concluded that PWN tail direction is not a good indicator of the pulsar's proper motion. The angle of tail deflection away from the puslar's proper motion is heavily influenced by the local flow field of the SNR ejecta, and the degree of this deflection may be used to reveal properties of the remnant itself.

The method of PWN inflation during subsonic expansion is dependent on events during the reverse shock interaction phase. For a pulsar that does not separate from its PWN, the PWN continues to inflate normally; for a pulsar which does partially separate and remains linked to its PWN through a tail, the fresh gas travels down the tail and inflates the PWN; for a pulsar which separates from the PWN entirely, the relic remains approximately constant in size while a secondary PWN inflates around the pulsar. As these events are responsible for much of the asymmetry seen in our model, we have determined the main parameters driving structure are ISM uniformity and total pulsar spin-down energy, with secondary contributions from factors such as pulsar trajectory $\psi_\text{kick}$ and initial spin-down luminosity. \\


Computer simulations were run at the Texas Advanced Computing Center (TACC) at The University of Texas at Austin using an allocation from the Extreme Science and Engineering Discovery Environment (XSEDE), which is supported by National Science Foundation grant number ACI-1053575.

POS acknowledges partial support from NASA Contract NAS8-03060. This research was carried out with partial support from Chandra Grant TM6-17002X. The authors would also like to thank the Physics Department at the University of Vermont for their hospitality for a research visit during which portions of this work was completed.

\bibliography{references}

\begin{thebibliography}{}
\expandafter\ifx\csname natexlab\endcsname\relax\def\natexlab#1{#1}\fi

\bibitem[{{Blondin} {et~al.}(2001){Blondin}, {Chevalier}, \&
  {Frierson}}]{2001ApJBlondin}
{Blondin}, J.~M., {Chevalier}, R.~A., \& {Frierson}, D.~M. 2001, ApJ, 563, 806

\bibitem[{{Borkowski} {et~al.}(1997){Borkowski}, {Blondin}, \&
  {McCray}}]{1997ApJBorkowski}
{Borkowski}, K.~J., {Blondin}, J.~M., \& {McCray}, R. 1997, ApJ, 477, 281

\bibitem[{{Borkowski} {et~al.}(2014){Borkowski}, {Reynolds}, {Green}, {Hwang},
  {Petre}, {Krishnamurthy}, \& {Willett}}]{2014ApJLBorkowski}
{Borkowski}, K.~J., {Reynolds}, S.~P., {Green}, D.~A., {et~al.} 2014, ApJL,
  709, L18

\bibitem[{{Bucciantini} {et~al.}(2004){Bucciantini}, {Amato}, {Bandiera},
  {Blondin}, \& {Del Zanna}}]{2004A&ABucciantini}
{Bucciantini}, N., {Amato}, E., {Bandiera}, R., {Blondin}, J.~M., \& {Del
  Zanna}, L. 2004, A\&A, 423, 253

\bibitem[{{Chevalier}(1982)}]{1982ApJChevalier}
{Chevalier}, R.~A. 1982, ApJ, 258, 790

\bibitem[{{Colella} \& {Woodward}(1984)}]{1984JCPC&W}
{Colella}, P., \& {Woodward}, P.~R. 1984, J. Comput. Phys., 54, 174

\bibitem[{{Ferriera} \& {de Jager}(2008)}]{2008A&AFerreira}
{Ferriera}, S. E.~S., \& {de Jager}, O.~C. 2008, A\&A, 478, 17

\bibitem[{{Gelfand} {et~al.}(2009){Gelfand}, {Slane}, \&
  {Weiqun}}]{2009ApJGelfand}
{Gelfand}, J.~D., {Slane}, P.~O., \& {Weiqun}, Z. 2009, ApJ, 703, 2051

\bibitem[{{Hales} {et~al.}(2009){Hales}, {Gaensler}, {Chatterjee}, {van der
  Swaluw}, \& {Camilo}}]{2009ApJHales}
{Hales}, C., {Gaensler}, B., {Chatterjee}, S., {van der Swaluw}, E., \&
  {Camilo}, F. 2009, ApJ, 706, 1316

\bibitem[{{Hnatyk} \& {Petruk}(1999)}]{1999A&AHnatyk}
{Hnatyk}, B., \& {Petruk}, O. 1999, A\&A, 344, 295

\bibitem[{{Jun} \& {Jones}(1999)}]{1999ApJJun}
{Jun}, B., \& {Jones}, T.~W. 1999, ApJ, 511, 774

\bibitem[{{Kaspi} {et~al.}(2001){Kaspi}, {Roberts}, {Vasisht}, {Gotthelf},
  {Pivovaroff}, \& {Kawai}}]{2001ApJKaspi}
{Kaspi}, V.~M., {Roberts}, M.~E., {Vasisht}, G., {et~al.} 2001, ApJ, 560, 371

\bibitem[{{McKee} \& {Truelove}(1995)}]{1995PyRepMcKee}
{McKee}, C.~F., \& {Truelove}, J.~K. 1995, Phys. Rep., 256, 157

\bibitem[{{Park} {et~al.}(2007){Park}, {Hughes}, {Slane}, {Burrows},
  {Gaensler}, \& {Ghavamian}}]{2007ApJPark}
{Park}, S., {Hughes}, J., {Slane}, P., {et~al.} 2007, ApJ, 670, L121

\bibitem[{{Reynolds} \& {Chevalier}(1984)}]{1984ApJReynolds}
{Reynolds}, S.~P., \& {Chevalier}, R.~A. 1984, ApJ, 278, 630

\bibitem[{{Slane} {et~al.}(2000){Slane}, {Chen}, {Schulz}, \&
  {Seward}}]{2000ApJSlane}
{Slane}, P., {Chen}, Y., {Schulz}, N.~S., \& {Seward}, F.~D. 2000, ApJL, 533,
  L29

\bibitem[{{Slane} {et~al.}(2008){Slane}, {Helfand}, {Reynolds}, {Gaensler},
  {Lemiere}, \& {Wang}}]{2008ApJLSlane}
{Slane}, P., {Helfand}, J.~D., {Reynolds}, S.~P., {et~al.} 2008, ApJL, 676, 33

\bibitem[{{Slane} {et~al.}(2012){Slane}, {Hughes}, {Temim}, {Rousseau},
  {Castro}, {Foight}, {Gaensler}, {Funk}, {Lemoine-Goumard}, {Gelfand},
  {Moffett}, {Dodson}, \& {Bernstein}}]{2012ApJSlane}
{Slane}, P., {Hughes}, J., {Temim}, T., {et~al.} 2012, ApJ, 749, 131

\bibitem[{{Temim} {et~al.}(2013){Temim}, {Slane}, {Castro}, {Plucinsky},
  {Gelfand}, \& {Dickel}}]{2013ApJTemim}
{Temim}, T., {Slane}, P., {Castro}, D., {et~al.} 2013, ApJ, 768, 61

\bibitem[{{Temim} {et~al.}(2015){Temim}, {Slane}, {Kolb}, {Blondin}, {Hughes},
  \& {Bucciantini}}]{2015ApJTemim}
{Temim}, T., {Slane}, P., {Kolb}, C., {et~al.} 2015, ApJ, 808, 100

\bibitem[{{van der Swaluw} {et~al.}(2001){van der Swaluw}, {Achterberg},
  {Gallant}, \& {T{\'o}th}}]{2001A&AVDS}
{van der Swaluw}, E., {Achterberg}, A., {Gallant}, Y.~A., \& {T{\'o}th}, G.
  2001, A\&A, 380, 309

\bibitem[{{van der Swaluw} {et~al.}(2004){van der Swaluw}, {Downes}, \&
  {Keegan}}]{2004A&AVDS}
{van der Swaluw}, E., {Downes}, T.~P., \& {Keegan}, R. 2004, A\&A, 420, 937

\bibitem[{{Wang} \& {Gotthelf}(1998)}]{1998ApJWang}
{Wang}, Q.~D., \& {Gotthelf}, E.~V. 1998, ApJL, 509, L109

\bibitem[{{Warren} \& {Blondin}(2013)}]{2013MNRASWarren}
{Warren}, D.~C., \& {Blondin}, J.~M. 2013, MNRAS, 429, 3099

\bibitem[{{Wongwathanarat} {et~al.}(2010){Wongwathanarat}, {Hammer}, \&
  {M{\"u}ller}}]{2010A&AWong}
{Wongwathanarat}, A., {Hammer}, N.~J., \& {M{\"u}ller}, E. 2010, A\&A, 560, A48

\end{thebibliography}

\end{document}